\documentclass[3p]{elsarticle}
\usepackage{graphicx}
\usepackage{envmath}
\usepackage{amssymb}
\usepackage{enumerate}
\usepackage{xcolor}
\usepackage[normalem]{ulem}
\usepackage{amsmath}
\usepackage{amsthm}

\newcommand{\ce}{{\mathcal C}}

\theoremstyle{definition}
\newtheorem{definition}{Definition}[section]

\begin{document}

\begin{frontmatter}

\title{Partially implicit Runge-Kutta methods for wave-like equations}

\author[DMA]{I. Cordero-Carri\'on}
\author[DAA]{P. Cerd\'a-Dur\'an}

\address[DMA]{Departamento de Matem\'aticas, University of Valencia, 
					 C/ Dr. Moliner, 50, E-46100 Burjassot, Spain }
\address[DAA]{Departamento de Astronom\'ia y Astrof\'isica, University of Valencia,
					 C/ Dr. Moliner, 50, E-46100 Burjassot, Spain}

\date{Received: date / Accepted: date}

\begin{abstract}
In this work we present a new class of Runge-Kutta (RK) methods for solving 
systems of hyperbolic equations with a particular structure, generalization of 
a wave-equation. The new methods are {\it partially implicit} in the sense that
a proper subset of the equations of the system contains some terms which are 
treated implicitly. These methods can be viewed as a particular case of the 
implicit-explicit (IMEX) RK methods for systems of equations with wave-like 
structure. For these systems, the optimal methods with the new structure are 
easier to derive than the IMEX ones, specially when aiming at higher-order (up 
to fourth-order in this work). The methods are constructed considering the 
classical strong-stability-preserving optimal explicit RK methods for the 
purely explicit part. The resulting partially implicit RK methods do not 
require any inversion of operators and hence their computational cost per 
iteration is similar to those of explicit RK methods. We analyse the stability 
and convergence properties and show their practical applicability in several 
numerical examples. Our results show that, compared with explicit RK methods, 
the new methods have better stability properties (larger steps are allowed) and
in general show smaller discretization error.
\end{abstract}

\begin{keyword}
ODEs; Runge-Kutta methods; IMEX; wave equation.
\end{keyword}

\end{frontmatter}

\section{Introduction}
The evolution in time of many complex systems, governed by partial differential
equations (PDE), implies, in a broad variety of cases, looking for the 
numerical solution of a system of ordinary differential equations (ODEs). The 
most commonly used methods to integrate in time these systems of ODEs are the 
well-known Runge-Kutta (RK) schemes (see e.g. \cite{Butcher} for a general 
review of these methods and their main properties). Several classifications of 
the RK methods can be done, according to, e.g., their convergence order, the 
number of stages or their explicit/implicit structure. The numerical stability 
of different numerical methods depends on the particular structure of the 
equations being solved. Therefore, it is necessary to choose the appropriate 
numerical scheme tailored to the particular necessities of the problem being 
solved.

The presence of stiff terms in systems of PDEs can lead to numerical 
instabilities if explicit RK (ERK, hereafter) schemes are used. A possible 
solution is the use of an implicit or partially implicit method in order to 
have numerically stable solutions without large constraints in the step size of
the discretisation. As an example, the (implicit) backward Euler method is 
numerically stable in cases where the (explicit) forward Euler method would 
fail (see e.g. \cite{Butcher}). As a second example, the so-called 
implicit-explicit (IMEX) RK methods have been used to solve 
convection-diffusion-reaction equations \cite{ARW,ARS,Pareshi,ParRus}, in 
situations where explicit methods would have failed. However, there are 
problems which, despite of not being stiff, still could benefit of the implicit
character of a numerical method. In this work we refer to stiff problems as 
those in which there is a large separation of scales. In these kind of problems
explicit methods fail when the step size is adapted to the slowest varying 
scale but not to the fastest one (see examples in \cite{Hairer2}). However, in 
a more general sense stiffness can be defined as those cases in which explicit 
methods fail \cite{Curtiss}. With the latter definition, the previous example 
could be called stiff. Further discussion about the different definitions of 
stiffness can be found in e.g. \cite{Soderlind}.

Let us consider, as an example, the harmonic oscillator
\begin{equation}
u_{tt} + \omega^2 u = 0, 
\end{equation}
$\omega$ being a real constant. If we define $v \equiv u_t$, this equation can 
be written as a system of first-order ODEs. In matrix form the system reads:
\begin{equation}
	U_t = A U \quad ; \quad 
U \equiv \begin{pmatrix} u \\ v \end{pmatrix} \quad ; \quad
A \equiv \begin{pmatrix} 0 & 1 \\ -\omega^2 & 0 \end{pmatrix}.
\end{equation}
Let us consider three different first-order integration methods:
\begin{align}
U_n&= U_{n-1} + h \, A \, U_{n-1} , & \text {(Euler method)} \\
U_n&= U_{n-1} + h \, A \, U_{n}, & \text {(implicit Euler method)} \\
U_n&= U_{n-1} + h \begin{pmatrix} 0 & 1 \\ 0&0\end{pmatrix} U_{n-1}
+ h \begin{pmatrix} 0 & 0 \\ -\omega^2&0\end{pmatrix} U_{n},
&  \text {(semi-implicit Euler method)}
\end{align}
being $h$ a fixed step size. Note that the semi-implicit Euler method is also 
known as symplectic Euler \cite{deVogelaere} due to its symplectic properties.
A necessary condition for the stability of the method (A-stability, see 
e.g.~\cite{Butcher}) is that the eigenvalues of the iteration matrix, $e$, are 
such that $|e|\le 1$. For the three methods described above these are given by 
(see e.g. \cite{Hairer}, Sect. I.5.2):
\begin{align}
	e = 1 + h \sigma, & \quad \text {(Euler method)} \\
	e = (1-h \sigma)^{-1}, & \quad \text {(implicit Euler method)} \\
	e^2 - (2 + h^2\sigma^2)e + 1 = 0 ,& \quad \text{(semi-implicit Euler method)}
\end{align}
being $\sigma$ the eigenvalues of the matrix $A$, which in this example are 
$\sigma = \pm i\,\omega$. As a result, the explicit Euler method is unstable 
for any value of $h>0$, while the implicit Euler method does not suffer from 
this instability for any $h\ge 0$. The semi-implicit Euler is stable for 
$|h\omega|<2$. This simple system is a clear example of a case in which, 
despite of not being stiff, an explicit method fails, while an implicit or 
semi-implicit method solves this numerical instability problem, at least for 
reasonable values of $h$. It seems natural to question whether we can identify 
a class of non-stiff problems which suffer from numerical instabilities if 
solved using explicit methods, and whether this instabilities can be cured by 
adding some implicit treatment in the numerical method. 

In this work we focus on systems of ODEs describing oscillatory behaviour or, 
in the case of PDEs, describing wave-like behaviour. A proper definition is 
given in the next section. Although this class of systems is not completely 
general, it is relevant for a large class of problems in many fields of 
mathematics, physics and engineering. Examples of applications in which it is 
necessary to solve systems of ODEs resulting in highly oscillatory solutions 
are: calculation of periodic orbits (see e.g. discussion in \cite{Hairer}); 
dynamics of biological systems, e.g. Lotka-Volterra equations for the modelling
of predator-prey systems (see e.g. \cite{Butcher}); electromagnetic transients 
in circuit networks (see e.g. \cite{Vlach}); molecular dynamics simulations 
(see e.g. \cite{Rapaport}) and computer animation (see e.g. \cite{Baraff}), 
among others. Examples of applications in which it is necessary to solve wave 
equations for either scalar or tensional variables are: Klein-Gordon equations,
with applications in solid state physics, nonlinear optics and quantum field 
theory; Maxwell equations, with applications in antenna design, medical 
imaging, photonics and geophysics (ground penetrating radar), e.g. FDTD methods
\cite{Yee}; Einstein equations describing gravitational radiation, with 
applications in gravitational wave astronomy; Navier-Cauchy equations for 
linear elasticity, with applications in seismology, among others. Additionally,
in fluid dynamics one can found multiple cases of systems of equations which 
admit solutions in form of waves and hence fall into the class of systems under
study. Examples of such systems are the Navier-Stokes equations for 
compressible fluids and shallow water equations, with interest in oceanography,
among others. 

In most of the cases above, examples in the literature can be found where 
explicit method show unstable behaviour or limit in excess the step-size. In 
those cases the customary solution is using a numerical method with some degree
of implicitness. For systems of ODEs the semi-implicit (symplectic) Euler 
method is known to have better stability properties than the explicit Euler one
for many of the examples above (see discussion on orbits and Lotka-Volterra 
equations in \cite{Butcher,Hairer} and an example for electromagnetic 
transients in \cite{Niiranen}). For cases involving the integration of 
trajectories (orbits, molecular dynamics, computer animation) the Verlet 
algorithm (also its variant velocity-Verlet) and leapfrog method are commonly 
used. Note that all these methods (semi-implicit Euler, Verlet, leapfrog) are 
very similar in their construction, they have symplectic properties and they 
are stable for systems with imaginary eigenvalues (as in the case of 
oscillatory systems). One important feature of these methods is that they are 
applicable to systems of ODEs in which one of the variables is updated first 
and the updated value is used to update the second variable. So despite of 
being explicit methods, they contain some degree of implicitness. We refer to 
methods sharing the latter property as {\it partially implicit}, which we will 
define more precisely in the next sections. Some of these instability problems 
are partially solved when using higher-order explicit Runge-Kutta (RK) schemes 
(as the popular fourth-order RK, see \cite{Asher} for their stability 
properties). Alternatively explicit (e.g. Adams-Bashford) or implicit 
(Adams-Moulton) linear multi-step methods improve the stability properties, 
albeit at the expense in increasing the complexity of the algorithm 
\cite{Butcher}. For systems with a Hamiltonian description (e.g. orbits) 
higher-order symplectic methods have been used traditionally to ensure 
stability and high accuracy (see e.g. \cite{Hairer}). For wave-like equations 
it is also possible to find cases where an improvement over the explicit Euler 
method has been reported when adding some degree of implicitness: the FDTD 
method to solve Maxwell equations uses a staggered method (partially implicit 
in our notation) to prevent numerical instabilities \cite{Yee}. There have been
attempts to improve the stability of the original FDTD method using the 
semi-implicit Euler method \cite{Fang} or fully implicit ones \cite{Holland}. 
Similarly, \cite{Moxley} proposed a staggered method, similar to FDTD, to solve
Schr\"odinger equations. In the case of Einstein equations the numerical 
instabilities in the first simulations of black hole mergers were overcome by a
variant of the Verlet method \cite{Pretorius} or using the partially implicit 
Crank-Nicholson method (see e.g. \cite{Baker}); nowadays, the most popular time
integrators for the Einstein equations are explicit high-order Runge-Kuta 
methods (fourth or higher-order, see \cite{Centrella}), although some amount of
Kreiss-Olinger numerical dissipation \cite{KreOli} is usually necessary to have
numerical stability \cite{Centrella}.

The goal of this work is to derive a set of numerical methods which are an 
extension of the semi-implicit Euler method to higher-orders in the sense that 
they are partially implicit (but without its symplectic behaviour). We present 
partially implicit RK (PIRK) methods of up to fourth-order which are applicable
to wave-like equations. These new PIRK methods can be viewed as a particular 
case of the IMEX RK methods applied to systems of equations with a particular 
structure (wave-like), but their derivation, as shown in this paper, is much 
simpler, allowing us to derive relatively easy methods up to fourth-order of 
convergence. The PIRK methods are able to provide stable evolutions due to 
their implicit component, but they do not require any analytical or numerical 
inversion of operators, so their computational costs are similar to those of 
the ERK methods. This provides a great advantage with respect to implicit or 
IMEX methods which need, in general, an inversion of some operators; depending 
on the complexity of the equations, the inversion can be done analytically or 
numerically, or be even prohibitive in practice from the numerical point of 
view. The choice of the coefficients in their derivation are based on stability
properties for both the explicit and implicit parts, as it is described in the 
following sections. The methods are validated by several numerical examples, 
for which stable evolutions are obtained, and compared with the results 
obtained by using ERK methods.

The manuscript is organized as follows. In Sect.~\ref{sect:eq-req} we introduce
the definition of {\it system of wave-like equations}. The structure of the 
PIRK methods, applied to this kind of equations is described in 
Sect.~\ref{sec:pirk}. In Sect.~\ref{sect:met-stab} we derive the PIRK methods 
up to fourth-order by analysing their stability properties. In 
Sects.~\ref{sect:num:ode} to \ref{sect:num:nl} we present numerical 
applications of the PIRK methods showing their stability and convergence 
properties. Conclusions are drawn in Sect.~\ref{sect:con}.

\section{Systems of wave-like equations}
\label{sect:eq-req}
Let us consider the following system of PDEs,
\begin{System}
u_t = \mathcal{L}_1 (u, v), \\
v_t = \mathcal{L}_2 (u) + \mathcal{L}_3 (u, v),
\label{e:system}
\end{System}
being $\mathcal{L}_1$, $\mathcal{L}_2$ and $\mathcal{L}_3$ general non-linear 
differential operators. Note that $\mathcal{L}_1$ and $\mathcal{L}_3$ depend on
both $u$ and $v$ variables, while $\mathcal{L}_2$ is only a function of $u$. 
Let us denote $(\bar{\alpha}_1 u, \bar{\alpha}_2 v)$, $\bar{\lambda} u$ and 
$(\bar{\gamma}_1 u, \bar{\gamma}_2 v)$ the associated linearized parts of the 
${\cal L}_1$, ${\cal L}_2$ and ${\cal L}_3$ operators with respect to the 
$(u, v)$ variables, respectively. We will consider that all these factors are 
real numbers. The linearized system can then be written as
\begin{System}
u_t = \bar{\alpha}_1 u + \bar{\alpha}_2 v, \\
v_t = \bar{\gamma}_1 u + \bar{\gamma}_2 v + \bar{\lambda} u,
\label{e:lin-system}
\end{System}
which in matrix form reads
\begin{equation}
	U_t = A U \quad ; \quad
U \equiv \begin{pmatrix} u \\ v \end{pmatrix} \quad ; \quad
A = \begin{pmatrix} \bar{\alpha}_1 & \bar{\alpha}_2 \\ 
\bar{\gamma}_1 + \bar{\lambda} & \bar{\gamma}_2 \end{pmatrix} .
\end{equation}
The eigenvalues of $A$ are
\begin{equation}
	\sigma_\pm = \frac{1}{2} \left [ \bar{\alpha}_1 + \bar{\gamma}_2 \pm \sqrt{
(\bar{\alpha}_1 - \bar{\gamma_2})^2 + 4 \bar{\alpha}_2 (\bar{\gamma_1} 
+ \bar{\lambda})}\right ].
\end{equation}
Given that the coefficients are real, wave-like solutions of these equations 
(or oscillatory solutions for the case of ODEs) appear if the eigenvalues have 
a non vanishing imaginary part.
\begin{definition}
The system (\ref{e:system}), whose linear part has the form 
(\ref{e:lin-system}), is said to be a system of {\it wave-like equations} if 
\begin{equation}
	(\bar{\alpha}_1 - \bar{\gamma_2})^2 + 4 \bar{\alpha}_2 (\bar{\gamma_1} 
+ \bar{\lambda}) < 0.
\end{equation}
\label{def:wave-like}
\end{definition}
A system of wave-like equations can be viewed as a generalisation of a 
second-order PDE wave equation when written as a first-order system in time. 
Note that, as a consequence of previous definition, 
$\bar{\alpha}_2 (\bar{\gamma}_1 + \bar{\lambda})<0$ is a necessary condition 
for the system to be of wave-like equations. 

Our aim is to deal with systems in which the terms responsible for the 
wave-like properties appear in the $\mathcal{L}_3$ operator. Keeping this in 
mind, we introduce the following definition:
\begin{definition}
The system (\ref{e:system}), whose linear part has the form 
(\ref{e:lin-system}), is said to be a system of {\it separable wave-like 
equations} if it is a system of {\it wave-like equations} (definition 
\ref{def:wave-like}) and $\bar{\alpha}_2 \bar{\lambda} < 0$.
\label{def:sep-wave-like}
\end{definition}
The adjective {\it separable} is used here to indicate that it is possible to 
separate the differential operator for $\partial_t v$ in such a way that 
$\bar{\alpha}_2 \bar{\lambda} < 0$. As we show in the numerical examples of 
Sects.~\ref{sect:num:ode} to \ref{sect:num:nl}, this condition is fairly 
easily fulfilled in typical examples of {\it wave-like equations}. The property
of {\it separable wave-like equations} is of help in the next sections for the 
stability analysis of the PIRK methods.

\section{Partially implicit Runge-Kutta methods}
\label{sec:pirk}

Let us consider a discretisation of the differential operators $\mathcal{L}_1$,
$\mathcal{L}_2$ and $\mathcal{L}_3$, appearing in a system of separable 
wave-like equations of the form (\ref{e:system}), which we denote by $L_1$, 
$L_2$ and $L_3$, respectively. No restrictions onto the discrete operators are 
imposed. Following the philosophy of the semi-implicit Euler method we want to 
build methods in which $L_1$ and $L_3$ are treated in an explicit way, whereas 
the $L_2$ operator is treated implicitly, and contains the terms responsible 
for the numerical instabilities.
\begin{definition}
We define a {\it partially implicit Runge Kutta} (PIRK) scheme as a numerical 
integration scheme in which:
\begin{enumerate}
\item The variable $u$ is evolved explicitly by means of a ERK scheme. Each 
stage of the ERK for $u$ corresponds to a stage of the PIRK scheme.
\item In each stage of the PIRK method, the variable $v$ is evolved taking into
account the updated value of $u$ for the evaluation of the $L_2$ operator.
\item The evaluation of $L_1$, $L_2$ and $L_3$ at each stage is performed at 
the same time level.
\item If $\mathcal{L}_2=0$ an ERK scheme is recovered. 
\end{enumerate}
\end{definition}
The general structure of a PIRK method to integrate $t$ from $t^n$ to 
$t^{n+1} = t^n + \Delta t$, being $\Delta t$ the time step, in $s$ stages is 
the following one: 
\begin{System}
u^{(0)} = u^{n}, \\
v^{(0)} = v^{n}, 
\end{System}
where $u^n$ and $v^n$ are the values of the variables at $t^n$; for the 
intermediate stages $i=1,\ldots,s-1$, intermediate variables can be computed as
\begin{System}
u^{(i)} = u^{n} + \Delta t \sum_{j=0}^{i-1} a_{i+1j+1} L_1 (t^{(j)}, u^{(j)}, v^{(j)}), \\
v^{(i)} = v^{n} + \Delta t \sum_{j=0}^{i} {\tilde a}_{i+1j+1} L_2 (t^{(j)}, u^{(j)}) 
+ \Delta t \sum_{j=0}^{i-1} a_{i+1j+1} L_3 (t^{(j)}, u^{(j)}, v^{(j)}),
\end{System}
where $t^{(j)} \equiv t^n + c_{j+1} \Delta t$; finally, the values of the 
variables at $t^{n+1}$ are 
\begin{System}
u^{n+1} = u^{(s)} = u^{n} + \Delta t \sum_{j=0}^{s-1} b_{j+1} L_1 (t^{(j)}, u^{(j)}, v^{(j)}), \\
v^{n+1} = v^{(s)}=v^{n} + \Delta t \sum_{j=0}^{s} {\tilde b}_{j+1} L_2 (t^{(j)}, u^{(j)})
+ \Delta t \sum_{j=0}^{s-1} b_{j+1} L_3 (t^{(j)}, u^{(j)}, v^{(j)}) ,
\end{System}
where the coefficients $a_{ij}$, ${\tilde a}_{ij}$, $b_{j}$ and ${\tilde b}_j$ 
are real constants and
\begin{equation}
	c_i = \sum_{j=1}^{s} a_{ij}.
\end{equation}
The coefficients $a_{ij}$, $b_{j}$ and $c_i$, with $i,j = 1, \dots, s$, 
correspond to the coefficients of a Butcher tableau (see e.g. \cite{Butcher}) 
for an ERK scheme with $s$ stages. These coefficients appear only in front of 
$L_1$ and $L_3$ operators. Therefore, if $L_2=0$ the method is equivalent to an
ERK. 

The coefficients ${\tilde a}_{ij}$ and ${\tilde b}_{j}$, in front of $L_2$ 
operator, do not correspond to a $s$-stages RK method because, in general, 
there is a non-zero ${\tilde b}_{s+1}$ coefficient. However, they correspond to
a diagonally implicit RK (DIRK, see e.g. \cite{Butcher}) method of $s+1$ 
stages, where ${\tilde a}_{s+1\,j} = {\tilde b}_j$ and ${\tilde c}_i = c_i$. 
The latter condition is necessary to fulfill the condition that all operators 
are computed at the same time level, $t^{(i)}$, when evaluated at a given stage
$i$. In general, a PIRK method of $s$ stages can be written as a particular 
case of an IMEX RK method with $s+1$ stages with the Butcher's tableau given in
table~\ref{tab:pirk}. Note that the explicit part is formally extended to $s+1$
stages by adding an additional trivial step in the $s+1$ row.

\begin{table}
\caption{Tableau for the explicit (left) and the implicit (right) part of a 
general PIRK scheme written as an IMEX-RK one.}
\centering
\begin{tabular}{cc} \\ \hline \hline \\
\begin{tabular}{c|cccccc}
$0$& $0$& $0$& $0$& $\cdots$& $0$& $0$ \\
$c_2$& $a_{21}$& $0$& $0$& $\cdots$& $0$& $0$ \\
$c_3$& $a_{31}$& $a_{32}$& $0$& $\cdots$& $0$& $0$ \\
$\vdots$& $\vdots$& $\vdots$& $ \ddots$& $\cdots$& $\vdots$& $\vdots$ \\
$c_s$& $a_{s\,1}$& $a_{s\,2}$& $\ldots$& $a_{s\,s-1}$& $0$& $0$ \\
$1$& $b_{1}$& $b_{2}$& $\ldots$& $b_{s-1}$& $b_{s}$& $0$ \\ \hline
   & $b_{1}$& $b_{2}$& $\ldots$& $b_{s-1}$& $b_{s}$& $0$
\end{tabular}
\hspace{0.2cm}
\begin{tabular}{c|cccccc}
$0$& $0$& $0$& $0$& $\cdots$& $0$& $0$ \\
$c_2$& ${\tilde a}_{21}$& ${\tilde a}_{22}$& $0$& $\cdots$& $0$& $0$ \\
$c_3$& ${\tilde a}_{31}$& ${\tilde a}_{32}$& ${\tilde a}_{33}$& $\cdots$& $0$& $0$ \\
$\vdots$& $\vdots$& $\vdots$& $\ddots$& $\cdots$& $\vdots$& $\vdots$ \\
$c_s$& ${\tilde a}_{s\,1}$& ${\tilde a}_{s\,2}$& $\ldots$& ${\tilde a}_{s\,s-1}$& ${\tilde a}_{s\,s}$& $0$ \\
$1$& ${\tilde b}_{1}$& ${\tilde b}_{2}$& $\ldots$& ${\tilde b}_{s-1}$& ${\tilde b}_{s}$& ${\tilde b}_{s+1}$ \\ \hline
   & ${\tilde b}_{1}$& ${\tilde b}_{2}$& $\ldots$& ${\tilde b}_{s-1}$& ${\tilde b}_{s}$& ${\tilde b}_{s+1}$
\end{tabular}
\\ \\ \hline \hline
\end{tabular}
\label{tab:pirk}
\end{table}

By using this structure it is possible to evaluate the $L_2$ operator at each 
stage using previously computed values of  $u^{(j)}$, without the need of any 
analytical or numerical inversion of operators. This strategy results in a 
numerical method which is explicit in this sense and hence comparable in 
computational cost to ERK schemes. Note that the PIRK methods can be trivially 
generalised to systems of $3$ or more equations by considering variables $u$ 
and $v$ as vectors of variables.

Although the method, viewed as a IMEX-RK scheme, has $s+1$ stages, in practice 
it is only necessary to compute each $L_i$ operator, $i=1,2,3$, $s$ times per 
iteration. The reason is that the computation of $L_2(u^{(0)})=L_2(u^{n})$ 
coincides with the value of $L_2(u^{(s)})$ at the previous iteration, so it is 
necessary to compute $s+1$ stages only in the first iteration. Therefore, 
we denote this method as a PIRK one with $s$ stages.

\section{Numerical methods and stability analysis}
\label{sect:met-stab}

In this section we derive the PIRK methods up to fourth-order. We construct 
methods of first, second and third-order with one, two and three stages, 
respectively. For the fourth-order scheme we use five stages. To fix all the 
coefficients of the PIRK methods described in the previous section, we follow 
the following procedure:
\begin{enumerate}[i)]
\item We choose the coefficients $a_{ij}$ and $b_{j}$ to recover the optimal 
SSP ERK method when implicitly treated parts are neglected (i.e. when 
$L_2 = 0$).
\item We choose the coefficients $\tilde{a}_{ij}$ and $\tilde{b}_{j}$ such that
the implicit part fulfills the conditions of a DIRK scheme. This step imposes 
restrictions over the possible values of the $\tilde{a}_{ij}$ and 
$\tilde{b}_{j}$ coefficients.
\item We choose the remaining free coefficients optimising the stability 
properties of the numerical scheme.
\end{enumerate}

\subsection{Explicit part}
Numerical methods based on a nonlinear stability requirement are very 
desirable. For this reason we choose strong stability preserving (SSP) methods 
(see e.g. \cite{GoShu98,GoShu01}) for the explicit part of our PIRK schemes. 
Forward Euler method corresponds to the first-order optimal SSP method. 
Gottlieb and Shu~\cite{GoShu98} proved that the classical second-order (Heun's)
method,
\begin{eqnarray}
	U^{(0)} &=& U^n, \nonumber \\
	U^{(1)} &=& U^n + \Delta t \, L (U^n), \nonumber \\
	U^{n+1} &=& \frac{1}{2} U^n + \frac{1}{2} U^{(1)}
+ \frac{1}{2} \Delta t \, L(U^{(1)}),
\label{e:2ERK}
\end{eqnarray}
is the optimal second-order two-stage SSP ERK method (ERK2 hereafter), and that
the third-order method due to Shu and Osher~\cite{ShuOsher},
\begin{eqnarray}
	U^{(0)} &=& U^n, \nonumber \\
	U^{(1)} &=& U^n + \Delta t \, L (U^n), \nonumber \\
	U^{(2)} &=& \frac{3}{4} U^n + \frac{1}{4} U^{(1)}
+ \frac{1}{4} \Delta t \, L(U^{(1)}), \nonumber \\
	U^{n+1} &=& \frac{1}{3} U^n + \frac{2}{3} U^{(2)}
+ \frac{2}{3} \Delta t \, L(U^{(2)}),
\label{e:3ERK}
\end{eqnarray}
is the optimal third-order three-stage SSP ERK method (ERK3 hereafter). The 
optimal adjective, for a given number of stages, refers to a maximization of 
the corresponding Courant-Friedrichs-Lewy (CFL) value (1 in both cases) and the
efficiency in the storage requirement. This property is crucial to capture the 
right behavior of the evolution of the system. The fourth-order five-stage 
optimal SSP ERK with positive coefficients was found in~\cite{Kraaijevanger} 
and again independently in~\cite{SpiteriRuuth02}, and guaranteed optimal 
\cite{Ruuth06} (ERK4 hereafter); it has the following form:
\begin{eqnarray}
	U^{(0)} &=& U^n, \nonumber \\
	U^{(1)} &=& U^n + 0.391752226571890 \, \Delta t \, L (U^n), \nonumber \\
	U^{(2)} &=& 0.444370493651235 \, U^n \nonumber \\
		&&+ 0.555629506348765 \, U^{(1)} 
+ 0.368410593050371 \, \Delta t \, L(U^{(1)}), \nonumber \\
	U^{(3)} &=& 0.620101851488403 \, U^n \nonumber \\ 
	&&+ 0.379898148511597 \, U^{(2)} 
+ 0.251891774271694 \, \Delta t \, L(U^{(2)}), \nonumber \\
	U^{(4)} &=& 0.178079954393132 \, U^n \nonumber \\
	&&+ 0.821920045606868 \, U^{(3)} 
+ 0.544974750228521 \, \Delta t \, L(U^{(3)}), \nonumber \\
	U^{n+1} &=& 0.517231671970585 \, U^{(2)} \nonumber \\
	&&+ 0.096059710526147 \, U^{(3)} 
+ 0.063692468666290 \, \Delta t \, L(U^{(3)}) \nonumber \\
	&&+ 0.386708617503269 \, U^{(4)} 
+ 0.226007483236906 \, \Delta t \, L(U^{(4)}).
\label{e:4ERK}
\end{eqnarray}

We use these SSP ERK methods for the explicit part of the PIRK schemes. 
Therefore the stability properties of the PIRK methods, when $L_2=0$, are the 
same as those of the SSP methods, which guarantees a good stability properties 
of the explicit part. In the next sections we assume that the numerical method, 
applied to the system at hand, is stable when $L_2=0$.

Let us consider the linearized system~(\ref{e:lin-system}). The first step in 
all SSP ERK methods mentioned above for the explicit part of the system of 
equations (i.e. when $L_2 =0$ or equivalently $\bar{\lambda} = 0$) can be 
written as
\begin{equation}
\left(\begin{tabular}{c} $u$ \\ $v$ \end{tabular}\right)^{n+1} 
= \left(\begin{tabular}{cc}
      $1+\alpha_1$ & $\alpha_2$ \\
      $\gamma_1$ & $1+\gamma_2$
      \end{tabular}\right)
\left(\begin{tabular}{c} $u$ \\ $v$ \end{tabular}\right)^n,
\end{equation}
where $\alpha_i := \bar{\alpha}_i \Delta t$ and 
$\gamma_i := \bar{\gamma}_i \Delta t$. For completeness and later use we also 
define $\lambda := \bar{\lambda} \Delta t$. The matrix appearing in the 
previous equation is referred as the stability matrix. Let us denote 
$\omega_i$, $i=1,2$, its two eigenvalues. As we are assuming that the explicit 
part of the system is numerically stable, it is necessary that 
$|\omega_i| \leq 1$. Let us denote by dex and trex the determinant and trace of
the stability matrix of the explicit part, respectively. The condition 
$|\omega_i| \leq1 $ implies, in particular, that these quantities are bounded 
as
\begin{eqnarray}
	\mathrm{dex} = |\omega_1| |\omega_2| = 
|(1 + \alpha_1)(1 + \gamma_2) - \alpha_2 \gamma_1| \leq 1, \label{e:bound1} \\
	\mathrm{trex} = |\omega_1 + \omega_2| = 
|2 + \alpha_1 + \gamma_2| \leq 2.
\label{e:bound2}
\end{eqnarray}
We use this property of the explicit part in the next sections to ensure the 
stability of the PIRK methods.

\subsection{Stability considerations}
We consider next the stability properties of the PIRK methods once the implicit
part is included. We focus here on the linear stability of the system; the 
analysis of the linear stability is the most simple case regarding the study of
the stability of a system of equations, but if a method does not verify even 
this criteria it is obviously not stable in general. In most cases, the linear 
part of the system is the dominant one and the results obtained in the analysis
of the linear stability are reproduced in the numerical simulations.

Since we are considering a system of equations and the characteristic 
structure of eigenvalues and eigenvectors of the explicit and implicit parts 
do not coincide necessarily, we have to consider the global structure of the 
system and the analysis has to involve the matrices which update the variables 
from one time step to the next one. 

Let us denote by $M_i$ the matrix which updates values of the variables for a 
$i$th-order PIRK method,
\begin{equation}
	\left(\begin{tabular}{c}
				$u^{n+1}$ \\
				$v^{n+1}$
	      \end{tabular}\right) = M_i \left(\begin{tabular}{c}
	      																 $u^n$ \\
	      																 $v^n$ 
	                                       \end{tabular}\right).
\label{e:defM}
\end{equation}
Linear stability thus requires that the absolute value of the two eigenvalues 
associated to the matrix $M_i$ are bounded by 1. However, in order to simplify 
the derivation of the PIRK methods, we are going to relax this condition on the
eigenvalues of the matrix $M_i$ by a bound on its determinant, 
$|\det(M_i)| \leq 1$. The idea is to restrict in a very effective way the 
possible range of values for the remaining coefficients associated to the 
implicit operator; moreover, in the numerical examples of the next sections, 
we will show that the optimal values for the coefficients are very close or 
equal to the ones found when the bound on the determinant is used. The 
different restrictions coming from both the bound of the determinant (boundary 
of the stability region) and the eigenvalues will be shown in some of the 
numerical experiments (see Sect.~\ref{sect:num:sph}).

Since we are considering systems of separable wave-like equations (see 
definition \ref{def:sep-wave-like}), we can assume that $\lambda \alpha_2 < 0$ 
(see numerical examples in the following sections). The combination 
$\lambda \alpha_2$ will appear in numerous places and basically counts for the 
freedom in the definition of the variable $v$ up to a constant factor, without 
changing the nature of the system of equations. This quantity is not 
necessarily very large, since we are not considering here that this source term
is stiff in this sense; we will consider in our analysis the cases 
$|\lambda \alpha_2| \ll 1$, $\lambda \alpha_2 \approx -1$ and 
$|\lambda \alpha_2| \gg 1$, in order to check the validity of the derived 
methods in all the possible ranges.

\subsection{First-order method}
\label{sec:PIRK1}
A one-stage first-order PIRK method for a separable wave-like equation can be 
written in general as
\begin{System}
	u^{n+1} = u^n + \Delta t \, L_1 (u^n, v^n), \\
	v^{n+1} = v^n + \Delta t \left[ (1-\ce_1) L_2(u^n) + \ce_1 L_2(u^{n+1}) 
+ L_3(u^n, v^n) \right],
\label{e:syst1}
\end{System}
where $\ce_1$, a real constant, is the only free parameter. This method is a 
particular case for the system~(\ref{e:system}) of the so-called IMEX-$\theta$ 
method (see e.g. \cite{HV}), where $\ce_1$ is the $\theta$ parameter. According
to \cite{HV}, $\ce_1$ must be chosen such that $\ce_1 \geq 1/2$.

System~(\ref{e:syst1}) can be written as
\begin{equation}
	\left(\begin{tabular}{c}
				$u^{n+1}$ \\
				$v^{n+1}$
	      \end{tabular}\right) = M_1 \left(\begin{tabular}{c}
	      																 $u^n$ \\
	      																 $v^n$ 
	                                       \end{tabular}\right),
\end{equation}
where
\begin{equation}
	M_1 = \left(
	\begin{tabular}{cc}
	$1+\alpha_1$ & $\alpha_2$ \\
	$\gamma_1 + \lambda(1 + \alpha_1 \ce_1)$ & $1 + \gamma_2 + \lambda \alpha_2 \ce_1$
	\end{tabular}\right).
        \label{e:m1}
\end{equation}
Since $\det(M_1) = \mathrm{dex} - \lambda \alpha_2 (1-\ce_1)$ and taking into 
account the condition~(\ref{e:bound1}), the only value of $\ce_1$ which 
guarantees $|\det(M_1)| \leq 1 \, \forall (\lambda \alpha_2)$ is $\ce_1=1$. 
This value satisfies the condition $\ce_1 \geq 1/2$. The resulting method is:
\begin{System}
	u^{n+1} = u^n + \Delta t \, L_1 (u^n, v^n), \\
	v^{n+1} = v^n + \Delta t \left[ L_2(u^{n+1}) + L_3(u^n, v^n) \right].
\end{System}
This method, which we call PIRK1 hereafter, is well known in the literature as 
the ARS(1,1,1) method~\cite{ARS,ARW}. For the case $L_3 =0$ it coincides with 
the semi-implicit Euler method. Table~\ref{tab:pirk1} shows the tableau of the 
PIRK1 method expressed as an 2 stages IMEX-RK one. The derivation of the PIRK1 
scheme is presented here as an illustration of the procedure we want to follow 
to derive higher-order PIRK methods.

\begin{table}
\caption{Tableau for the explicit (left) and implicit (right) part of a general 
one-stage first-order PIRK method, expressed as an 2 stages IMEX-RK scheme. 
Values of $\ce_1$ for different schemes are also given.}
\centering
\begin{tabular}{cc} \\ \hline \hline \\
\begin{tabular}{c|cc}
0 & 0 & 0 \\
1 & 1 & 0 \\ \hline
  & 1 & 0
\end{tabular}
\hspace{0.5 cm}
\begin{tabular}{c|cc}
0 & 0 & 0 \\
1 & $1-\ce_1$ & $\ce_1$ \\ \hline
  & $1-\ce_1$ & $\ce_1$
\end{tabular} \\ \\ \hline \hline 
\end{tabular} \\
PIRK1: $\ce_1$ = 1. \\
ERK1 (forward Euler): $\ce_1$ = 0.
\label{tab:pirk1}
\end{table}
%

\subsection{Second-order method}
\label{sec:PIRK2}
Next we derive a two-stages second-order PIRK method (PIRK2 hereafter), which 
can be regarded as an IMEX-RK scheme with three-stages. We impose the 
two-stages second-order SSP optimal method for the purely explicit parts ($L1$ 
and $L3$ operators). The remaining coefficients are restricted by the order 
conditions (see \cite{KenCar} for second-order and $s=3$) and 
$c_i={\tilde c}_i$ (but $b_i \neq {\tilde b}_i$ in general). With these 
conditions, the PIRK2 method can be written in terms of two free real 
coefficients, $(\ce_1, \ce_2)$, as follows:
\begin{System}
	u^{(1)} = u^n + \Delta t \, L_1 (u^n, v^n), \\
	v^{(1)} = v^n + \Delta t \left[ (1-\ce_1) L_2(u^n) + \ce_1 L_2(u^{(1)}) 
+ L_3(u^n, v^n) \right],
\label{e:syst2a}
\end{System}
\begin{System}
	u^{n+1} = \frac{1}{2} \left[ u^n + u^{(1)} 
+ \Delta t \, L_1 (u^{(1)}, v^{(1)}) \right], \\
	v^{n+1} = v^n + \frac{\Delta t}{2} \left[ 
L_2(u^n) + 2 \ce_2 L_2(u^{(1)}) + (1 - 2 \ce_2) L_2(u^{n+1}) \right. \\
	\left. \hspace{2.6cm} + L_3(u^n, v^n) + L_3 (u^{(1)}, v^{(1)}) \right].
\label{e:syst2b}
\end{System}
From~(\ref{e:syst2a}) and (\ref{e:syst2b}),
\begin{equation}
	\left(\begin{tabular}{c}
				$u^{n+1}$ \\
				$v^{n+1}$
	      \end{tabular}\right) = M_2 \left(\begin{tabular}{c}
	      																 $u^n$ \\
	      																 $v^n$ 
	                                       \end{tabular}\right),
\end{equation}
where
\begin{eqnarray}
	M_2&=&\left( \begin{tabular}{cc}
	             1 & 0 \\
	             $(1/2-\ce_2) \lambda$ & 1
	             \end{tabular}\right) 
\left[ \left(
\begin{tabular}{cc}
1/2 & 0 \\
$(\lambda + \gamma_1)/2$ & $1 + \gamma_2/2$
\end{tabular}\right) \right. \nonumber \\
&&\left.+\frac{1}{2} \left( \begin{tabular}{cc}
	                          $1 + \alpha_1$ & $\alpha_2$ \\
	                          $\gamma_1 + 2 \lambda \ce_2$ & $\gamma_2$
	                          \end{tabular}\right)
\left( \begin{tabular}{cc}
       $1 + \alpha_1$ & $\alpha_2$ \\
       $\gamma_1+\lambda(1+\alpha_1 \ce_1)$ & $1+\gamma_2+\lambda \alpha_2 \ce_1$
	     \end{tabular}\right) \right]. \hspace{0.6cm}
\end{eqnarray}

Its determinant is given by:
\begin{eqnarray}
	\det(M_2) &=& \frac{1}{4}\left[(1-\mathrm{dex})^2 + \mathrm{trex}^2 
+ \lambda \alpha_2 (1-\mathrm{dex}) (1-2\ce_1+2\ce_2) \right. \nonumber \\
	&&\hspace{0.3cm} \left. + (\lambda \alpha_2)^2 (2\ce_2-\ce_1-2\ce_1\ce_2) \right].
\end{eqnarray}
The conditions $1-2\ce_1+2\ce_2 = 2\ce_2-\ce_1-2\ce_1\ce_2 =0$ guarantee 
$|\det(M_2)| \leq 1 \, \forall (\lambda \alpha_2)$, but they lead to 
$\ce_1=(1\pm i)/2 \notin \mathbb{R}$, $\ce_2=\pm i/2 \notin \mathbb{R}$, where 
$i = \sqrt{-1}$ is the imaginary unit. The condition on the bound for the 
determinant of $M_2$, $|\det(M_2)| \leq 1$, is equivalent to:
\begin{equation}
	-4 \leq K_1 + \lambda \alpha_2 \, K_2 (1-2\ce_1+2\ce_2) 
+ (\lambda \alpha_2)^2 (2\ce_2-\ce_1-2\ce_1\ce_2) \leq 4,
\end{equation}
where
\begin{equation}
	K_1 := (1-\mathrm{dex})^2+\mathrm{trex}^2 = 
1 + \omega_1^2 \omega_2^2 + \omega_1^2 + \omega_2^2 \in [1,4]
\label{e:bound3}
\end{equation}
and
\begin{equation}
	K_2 := 1-\mathrm{dex} \in [0,2].
\end{equation}
The minimum value of $K_1$ is reached for $\omega_1 = \omega_2 = 0$. The 
maximum value of $K_1$ is reached for $\omega_1^2 = \omega_2^2 = 1$. The 
minimum value of $K_2$ is reached for $\omega_1 = \omega_2 = \pm 1$. The 
maximum value of $K_2$ is reached for $\omega_1 = - \omega_2 = \pm 1$. For 
$\omega_1 = \omega_2 = 0$, $K_2 = 1$. We analyse the value of $\det(M_2)$ for 
$\omega_i = 0, \pm 1$, $i=1,2$, and in the cases $|\lambda \alpha_2| \ll 1$, 
$\lambda \alpha_2 \approx -1$ and $|\lambda \alpha_2| \gg 1$. The resulting 
sufficient conditions are (see~\ref{a:PIRK2} for more details): 
\begin{eqnarray}
	0 \leq 2\ce_2(1-\ce_1)-\ce_1, \; 1-2\ce_1+2\ce_2 \leq 0, \nonumber \\
	0 \leq 6+5\ce_1-6\ce_2+2\ce_1\ce_2, \; 0 \leq 4+\ce_1-2\ce_1\ce_2,
\label{e:PIRK2boundc}
\end{eqnarray}
which involve only coefficients $\ce_i$, together with
\begin{equation}
	-4 \leq \lambda \alpha_2 (1-2\ce_1+2\ce_2), \;
-5 \leq (\lambda \alpha_2)^2 (2\ce_2-\ce_1-2\ce_1\ce_2).
\label{e:PIRK2boundx}	
\end{equation}

For $|\lambda \alpha_2| \ll 1$, the first inequality in~(\ref{e:PIRK2boundx}) 
is the most relevant one. We impose $1-2\ce_1+2\ce_2=0$, and minimize 
$|2\ce_2(1-\ce_1)-\ce_1|$ (taking into account~(\ref{e:PIRK2boundc})). The 
resulting values for the coefficients are $\ce_1=1/2$ and $\ce_2=0$. The 
scheme, which we will call PIRK2a hereafter, is then written as:
\begin{System}
	u^{(1)} = u^n + \Delta t \, L_1 (u^n, v^n), \\
	v^{(1)} = v^n + \Delta t \left[\frac{1}{2} L_2(u^n) + \frac{1}{2} L_2(u^{(1)})
+ L_3(u^n, v^n) \right],
\end{System}
\begin{System}
	u^{n+1} = \frac{1}{2} \left[ u^n + u^{(1)} 
+ \Delta t \, L_1 (u^{(1)}, v^{(1)}) \right], \\
	v^{n+1} = v^n + \frac{\Delta t}{2} \left[ 
L_2(u^n) + L_2(u^{n+1}) + L_3(u^n, v^n)+ L_3 (u^{(1)}, v^{(1)}) \right].
\end{System}

For $|\lambda \alpha_2| \gg 1$, second inequality in~(\ref{e:PIRK2boundx}) is 
the most relevant one. We impose $2\ce_2-\ce_1-2\ce_1\ce_2=0$, and minimize 
$|1-2\ce_1+2\ce_2|$ (taking into account~(\ref{e:PIRK2boundc})). The resulting 
values for the coefficients are $\ce_1=1-\sqrt{2}/2$ and 
$\ce_2=(\sqrt{2}-1)/2$. The scheme, named PIRK2b hereafter, is written in this 
case as:
\begin{System}
	u^{(1)} = u^n + \Delta t \, L_1 (u^n, v^n), \\
	v^{(1)} = v^n + \Delta t \left[ \frac{\sqrt{2}}{2} L_2(u^n) 
+ \left(1 - \frac{\sqrt{2}}{2} \right) L_2(u^{(1)}) + L_3(u^n, v^n) \right],
\end{System}
\begin{System}
	u^{n+1} = \frac{1}{2} \left[ u^n + u^{(1)} 
+ \Delta t \, L_1 (u^{(1)}, v^{(1)}) \right], \\
	v^{n+1} = v^n + \frac{\Delta t}{2} \left[ 
L_2(u^n) + (\sqrt{2}-1) L_2(u^{(1)}) + (2-\sqrt{2}) L_2(u^{n+1}) \right. \\
	\left. \hspace{2.6cm} + L_3(u^n, v^n) + L_3 (u^{(1)}, v^{(1)}) \right].
\end{System}

Depending of the value of $|\lambda \alpha_2|$, it will be more convenient to 
use a particular set of values for the coefficients. This fact will be 
illustrated in numerical examples of Sects.~\ref{sect:num:ode} and 
\ref{sect:num:sph}. Table~\ref{tab:pirk2} shows the tableau of the second-order
PIRK method expressed as a 3 stages IMEX-RK scheme.

\begin{table}
\caption{Tableau for the explicit (left) and implicit (right) part of a general
2 stages second-order PIRK method, expressed as a 3 stages IMEX-RK scheme. 
Values of $\ce_1$ and $\ce_2$ for different particular schemes are also given.}
\centering
\begin{tabular}{cc} \\ \hline \hline \\
\begin{tabular}{c|ccc}
0 & 0 & 0 & 0 \\
1 & 1 & 0 & 0 \\ 
1 & $1/2$ & $1/2$ & 0 \\ \hline
  & $1/2$ & $1/2$ & 0
\end{tabular}
\hspace{0.5 cm}
\begin{tabular}{c|ccc}
0 & 0 & 0 & 0 \\
1 & $1-\ce_1$ & $\ce_1$ & 0 \\ 
1 & $1/2$ & $\ce_2$ & $1/2-\ce_2$ \\ \hline
  & $1/2$ & $\ce_2$ & $1/2-\ce_2$
\end{tabular} \\ \\ \hline \hline 
\end{tabular} \\
PIRK2a: $(\ce_1, \ce_2) = (1/2, 0)$ \\ 
PIRK2b: $(\ce_1, \ce_2) = (1-\sqrt{2}/2, (\sqrt{2} - 1)/2)$ \\
ERK2 (Heun's): $(\ce_1, \ce_2) = (0, 1/2)$
\label{tab:pirk2}
\end{table}

\subsection{Third-order method}
\label{sect:PIRK3}
The three-stages third-order PIRK method can be viewed as a four-stages 
IMEX-RK scheme. For the purely explicit parts ($L1$ and $L3$ operators) we 
impose the SSP optimal three-stages third-order method. The remaining 
coefficients are restricted by the order conditions (see \cite{KenCar} for 
third-order and $s=4$) and $c_i={\tilde c}_i$ (but $b_i \neq {\tilde b}_i$ in 
general). The resulting method can be written in terms of two real 
coefficients, $(\ce_1,\ce_2)$ (do not confuse these coefficients with the ones 
appearing in previous methods), as follows:
\begin{System}
	u^{(1)} = u^n + \Delta t \, L_1 (u^n, v^n), \\
	v^{(1)} = v^n + \Delta t \left[ (1-\ce_1) L_2(u^n) + \ce_1 L_2(u^{(1)}) 
+ L_3(u^n, v^n) \right],
\label{e:syst3a}
\end{System}
\begin{System}
	u^{(2)} = \frac{1}{4} \left[ 3 u^n + u^{(1)} 
+ \Delta t \, L_1 (u^{(1)}, v^{(1)}) \right], \\
	v^{(2)} = v^n + \frac{\Delta t}{4} \left[ 
2 (\ce_1 + 2 \ce_2) L_2(u^n) + 4 \ce_2 L_2(u^{(1)}) + 2(1 - \ce_1 - 4 \ce_2) L_2(u^{(2)})
\right. \\
	\left. \hspace{2.5cm} + L_3(u^n, v^n) + L_3 (u^{(1)}, v^{(1)}) \right],
\label{e:syst3b}
\end{System}
\begin{System}
	u^{n+1} = \frac{1}{3} \left[ u^n + 2 u^{(2)} 
+ 2 \Delta t \, L_1 (u^{(2)}, v^{(2)}) \right], \\
	v^{n+1} = v^n + \frac{\Delta t}{6} \left[ L_2(u^n) + L_2(u^{(1)}) 
+ 4 L_2(u^{(2)}) \right. \\
	\left. \hspace{2.7cm} + L_3(u^n, v^n) + L_3 (u^{(1)}, v^{(1)}) 
+ 4 L_3 (u^{(2)}, v^{(2)}) \right],
\label{e:syst3c}
\end{System}
From~(\ref{e:syst3a})-(\ref{e:syst3c}),
\begin{equation}
	\left(\begin{tabular}{c}
				$u^{n+1}$ \\
				$v^{n+1}$
	      \end{tabular}\right) = M_3 \left(\begin{tabular}{c}
	      																 $u^n$ \\
	      																 $v^n$ 
	                                       \end{tabular}\right),
\end{equation}
where
\begin{equation}
	M_3 = \left( \begin{tabular}{cc}
	             $1+\frac{\alpha_1}{6}$ & $\frac{\alpha_2}{6}$ \\
	             $\frac{\gamma_1+\lambda}{6}$ & $1+\frac{\gamma_2}{6}$
	             \end{tabular}\right) 
+ \left(\begin{tabular}{cc}
        $\alpha_1$ & $\alpha_2$ \\
        $\gamma_1+\lambda$ & $\gamma_2$
        \end{tabular}\right) \left( \frac{1}{6} N_1 + \frac{2}{3} N_2 \right),
\end{equation}
\begin{equation}
	N_1 = \left( \begin{tabular}{cc}
	             1 & 0 \\
	             $\lambda \ce_1$ & 1
	             \end{tabular}\right)
	      \left( \begin{tabular}{cc}
               $1+\alpha_1$ & $\alpha_2$ \\
               $\gamma_1+\lambda(1-\ce_1)$ & $1+\gamma_2$
               \end{tabular}\right),
\end{equation}
\begin{eqnarray}
	N_2 = \left( \begin{tabular}{cc}
	             1 & 0 \\
	             $(1-\ce_1-4\ce_2)\frac{\lambda}{2}$ & 1
	             \end{tabular}\right) \left[
\left(\begin{tabular}{cc}
      $1+\frac{\alpha_1}{4}$ & $\frac{\alpha_2}{4}$ \\
      $\frac{\gamma_1}{4}+(\ce_1+2\ce_2)\frac{\lambda}{2}$ & $1+\frac{\gamma_2}{4}$
      \end{tabular}\right) \right. \nonumber \\
\left. + \left( \begin{tabular}{cc}
	              $\frac{\alpha_1}{4}$ & $\frac{\alpha_2}{4}$ \\
	              $\frac{\gamma_1}{4}+\lambda \ce_2$ & $\frac{\gamma_2}{4}$
	              \end{tabular}\right) N_1 \right].
\end{eqnarray}

Its determinant is given by:
\begin{eqnarray}
	\det(M_3) &=& \frac{1}{36} \left[ 14 + 2(\mathrm{trex}-1)^3 
+ (\mathrm{dex}-2)^3 + 6 \, \mathrm{trex}^2
+ 3 \, \mathrm{dex}((\mathrm{trex}-1)^2-2) \right] \nonumber \\
	&&+ \frac{1}{24} \lambda \alpha_2 (-1 + \ce_1 - 4 \ce_2) \left[ (\mathrm{dex}-2)^2
+ (\mathrm{trex}-1)^2 -2 \right] \nonumber \\ 
	&&+ \frac{1}{12} (\lambda \alpha_2)^2 \left[ \ce_1 - 4 \ce_2 
+ (\mathrm{dex}-1)(4 \ce_2 - \ce_1^2 - 4 \ce_1 \ce_2) \right] \nonumber \\
	&&- \frac{1}{72} (\lambda \alpha_2)^3 \left[-1 + 3(1-2\ce_1)(\ce_1+4\ce_2) \right]. 
\end{eqnarray}

The expressions $-1 + \ce_1 - 4\ce_2$, $\ce_1 - 4\ce_2$, $4\ce_2 - \ce_1^2 - 4\ce_1 \ce_2$ and 
$-1 + 3(1 - 2\ce_1)(\ce_1 + 4\ce_2)$ cannot vanish simultaneously. The condition 
$|\det(M_3)| \leq 1$ is equivalent to:
\begin{eqnarray}
	-1 \leq &&\frac{K_3}{36} + \frac{\lambda \alpha_2 K_4 (-1 + \ce_1 - 4\ce_2)}{24} 
\nonumber \\
	&&+ \frac{(\lambda \alpha_2)^2}{12} 
[\ce_1 - 4\ce_2 + (\mathrm{dex}-1)(4\ce_2 - \ce_1^2 - 4\ce_1 \ce_2))] \nonumber \\
	&&- \frac{(\lambda \alpha_2)^3}{72} [-1 + 3(1 - 2\ce_1)(\ce_1 + 4\ce_2)] \leq 1,
\end{eqnarray}
where
\begin{equation}
	K_3 := 14 + 2(\mathrm{trex}-1)^3 + (\mathrm{dex}-2)^3 + 6\mathrm{trex}^2 
+ 3\mathrm{dex} [(\mathrm{trex}-1)^2 - 2] \in [-12,36],
\end{equation}
and
\begin{equation}
	K_4 := (\mathrm{dex}-2)^2 + (\mathrm{trex}-1)^2 - 2 \in [0,8].
\end{equation}
The minimum value of $K_3$ is reached for $\omega_1 = 1 = - \omega_2$ and 
$\omega_2 = 1 = -\omega_1$. For $\omega_1 = \omega_2 = -1$, $K_3 = 4$. The 
maximum value of $K_3$ is reached for $\omega_1 = \omega_2 = 1$. The minimum 
value of $K_4$ is reached for $\omega_1 = \omega_2 = 1$. The maximum value of 
$K_4$ is reached for $\omega_1 = \omega_2 = -1$, $\omega_1 = 1 = - \omega_2$ 
and $\omega_2 = 1 = - \omega_1$.

We analyse the bound for the determinant for the values $\omega_i = \pm 1$, 
$i=1,2$, and in the cases $|\lambda \alpha_2| \ll 1$, 
$\lambda \alpha_2 \approx -1$ and $|\lambda \alpha_2| \gg 1$. The resulting 
sufficient conditions are (see~\ref{a:PIRK3} for more details):
\begin{eqnarray}
	-\frac{20}{9} \leq \ce_1 - 4\ce_2 \leq 0, \;\; 
-1 + 3(1 - 2\ce_1)(\ce_1 + 4\ce_2) \leq 0, \nonumber \\
	0 \leq 73 + 18\ce_1^2 - 180\ce_2 + 9\ce_1(3 + 8\ce_2), \;\; 
0 \leq 9\ce_1 - 12\ce_2 - 6\ce_1^2 - 24\ce_1 \ce_2 + 143, \nonumber \\
	0 \leq 103 - 15\ce_1 - 6\ce_1^2 + 84\ce_2 - 24\ce_1 \ce_2, \;\; 
0 \leq 6\ce_1^2 - 15\ce_1 + 36\ce_2 + 24\ce_1 \ce_2 + 71, \nonumber \\
\label{e:PIRK3boundc}
\end{eqnarray}
which involve only coefficients $\ce_i$, together with
\begin{eqnarray}
	\lambda \alpha_2 (-1 + \ce_1 - 4\ce_2) \leq \frac{8}{3}, \;\; 
-24 \leq (\lambda \alpha_2)^2 (\ce_1 - 4\ce_2), \nonumber \\
	(\lambda \alpha_2)^3 [-1 + 3(1 - 2\ce_1)(\ce_1 + 4\ce_2)] \leq 48.
\label{e:PIRK3boundx}
\end{eqnarray}

For $|\lambda \alpha_2| \ll 1$, first inequality in~(\ref{e:PIRK3boundx}) is 
the most relevant one. Firstly, we minimize $|-1 + \ce_1 - 4\ce_2|$ (taking into 
account~(\ref{e:PIRK3boundc})). Consequently, we choose $\ce_2 = \ce_1/4$. This 
condition minimizes the factors accompanying $\lambda \alpha_2$ and 
$(\lambda \alpha_2)^2$ in~(\ref{e:PIRK3boundx}). The remaining inequalities 
from~(\ref{e:PIRK3boundc}) and (\ref{e:PIRK3boundx}) reduce to
\begin{equation}
	\frac{3 - \sqrt{1245}}{12} \leq \ce_1 \leq \frac{3 + \sqrt{1245}}{12}, \;\;
(\lambda \alpha_2)^3 (-1 + 6\ce_1 - 12\ce_1^2) \leq 48.
\end{equation}
Secondly, we minimize $|-1 + 6\ce_1 - 12\ce_1^2|$ (taking into account its allowed 
range). The minimum is placed at $\ce_1=1/4$. Therefore, the resulting values are
\begin{equation}
	(\ce_1, \ce_2) = \left( \frac{1}{4}, \frac{1}{16} \right), 
\end{equation}
and the method, which we call PIRK3a hereafter, is written as
\begin{System}
	u^{(1)} = u^n + \Delta t \, L_1 (u^n, v^n), \\
	v^{(1)} = v^n + \Delta t \left[\frac{3}{4} L_2(u^n) + \frac{1}{4} L_2(u^{(1)})
+ L_3(u^n, v^n) \right],
\end{System}
\begin{System}
	u^{(2)} = \frac{1}{4} \left[ 3 u^n + u^{(1)} 
+ \Delta t \, L_1 (u^{(1)}, v^{(1)}) \right], \\
	v^{(2)} = v^n + \frac{\Delta t}{4} \left[ \frac{3}{4} L_2(u^n) 
+ \frac{1}{4} L_2(u^{(1)}) + L_2(u^{(2)})
\right. \\ 
\left. \hspace{2.5cm} + L_3(u^n, v^n) + L_3 (u^{(1)}, v^{(1)}) \right],
\end{System}
\begin{System}
	u^{n+1} = \frac{1}{3} \left[ u^n + 2 u^{(2)} 
+ 2 \Delta t \, L_1 (u^{(2)}, v^{(2)}) \right], \\
	v^{n+1} = v^n + \frac{\Delta t}{6} \left[ L_2(u^n) + L_2(u^{(1)}) 
+ 4 L_2(u^{(2)}) \right. \\
	\left. \hspace{2.7cm} + L_3(u^n, v^n) + L_3 (u^{(1)}, v^{(1)}) 
+ 4 L_3 (u^{(2)}, v^{(2)}) \right].
\end{System}

For $|\lambda \alpha_2| \gg 1$, last inequality in~(\ref{e:PIRK3boundx}) is the
most relevant one. Firstly, we minimize $|-1 + 3(1 - 2\ce_1)(\ce_1 + 4\ce_2)|$
(taking into account~(\ref{e:PIRK3boundc})). Consequently, we choose 
$\displaystyle \ce_2 = \frac{1}{4}\left(\frac{1}{3(1-2\ce_1)} - \ce_1\right)$. This 
condition minimizes the factor accompanying $(\lambda \alpha_2)^3$ 
in~(\ref{e:PIRK3boundx}). The remaining inequalities from~(\ref{e:PIRK3boundc})
and (\ref{e:PIRK3boundx}) reduce to
\begin{equation}
	\frac{-17 - \sqrt{2377}}{72} \leq \ce_1 \leq \frac{-17 + \sqrt{2377}}{72},
\end{equation}
and
\begin{equation}
	\lambda \alpha_2 \left(-1 + \frac{(-1 + 6\ce_1 - 12\ce_1^2)}{3(2\ce_1-1)} \right) 
\leq \frac{8}{3}, \;\; 
24 \leq (\lambda \alpha_2)^2 \frac{(-1 + 6\ce_1 - 12\ce_1^2)}{3(2\ce_1-1)}.
\end{equation}
Secondly, we minimize 
$\displaystyle \left|\frac{(-1 + 6\ce_1 - 12\ce_1^2)}{3(2\ce_1-1)}\right|$ (taking 
into account its allowed range). The minimum is placed at 
$\displaystyle \ce_1 = \frac{3-\sqrt{3}}{6}$. Therefore, the resulting values are
\begin{equation}
	(\ce_1, \ce_2) = \left( \frac{3-\sqrt{3}}{6}, \frac{-1+\sqrt{3}}{8} \right), 
\end{equation}
and the method, which we call PIRK3b hereafter, is written as
\begin{System}
	u^{(1)} = u^n + \Delta t \, L_1 (u^n, v^n), \\
	v^{(1)} = v^n + \Delta t \left[ \frac{(3+\sqrt{3})}{6} L_2(u^n) 
+ \frac{(3-\sqrt{3})}{6} L_2(u^{(1)}) + L_3(u^n, v^n) \right],
\end{System}
\begin{System}
	u^{(2)} = \frac{1}{4} \left[ 3 u^n + u^{(1)} 
+ \Delta t \, L_1 (u^{(1)}, v^{(1)}) \right], \\
	v^{(2)} = v^n + \frac{\Delta t}{4} \left[ 
\frac{(3+\sqrt{3})}{6} L_2(u^n) + \frac{(-1+\sqrt{3})}{2} L_2(u^{(1)}) 
+ \frac{2}{3}(3-\sqrt{3}) L_2(u^{(2)})
\right. \\
	\left. \hspace{2.5cm} + L_3(u^n, v^n) + L_3 (u^{(1)}, v^{(1)}) \right],
\end{System}
\begin{System}
	u^{n+1} = \frac{1}{3} \left[ u^n + 2 u^{(2)} 
+ 2 \Delta t \, L_1 (u^{(2)}, v^{(2)}) \right], \\
	v^{n+1} = v^n + \frac{\Delta t}{6} \left[ L_2(u^n) + L_2(u^{(1)}) 
+ 4 L_2(u^{(2)}) \right. \\
	\left. \hspace{2.7cm} + L_3(u^n, v^n) + L_3 (u^{(1)}, v^{(1)}) 
+ 4 L_3 (u^{(2)}, v^{(2)}) \right].
\end{System}

As in the second-order PIRK methods, depending on the value of 
$|\lambda \alpha_2|$, it will be more convenient to use a particular set of 
values for the coefficients (see numerical examples in 
Sects.~\ref{sect:num:ode} and \ref{sect:num:sph}).

\begin{table}
\caption{Tableau for the explicit (left) and implicit (right) part of a general
three-stages third-order PIRK method, expressed as a four-stages IMEX-RK 
scheme. Values of $\ce_1$ and $\ce_2$ for different schemes are also given. 
Note that the tableau can be trivially simplified into 3 stages. $^*$With these
values this tableau is equivalent to the IMEX-SSP3(4,3,3) if applied to the 
system~(\ref{e:system}) (see section~\ref{sec:imexvspirk}).}
\centering
\begin{tabular}{cc} \\ \hline \hline \\
\begin{tabular}{c|cccc}
0 & 0 & 0 & 0 & 0 \\
1 & 1 & 0 & 0 & 0 \\ 
1/2 & $1/4$ & $1/4$ & 0 & 0 \\ 
1 & $1/6$ & $1/6$ & 2/3 & 0 \\ \hline
  & $1/6$ & $1/6$ & 2/3 & 0
\end{tabular}
\hspace{0.1 cm}
\begin{tabular}{c|cccc}
0 & 0 & 0 & 0 & 0 \\
1 & $1-\ce_1$ & $\ce_1$ &  0 & 0 \\ 
1/2 & $(\ce_1+2\ce_2)/2$ & $\ce_2$ & $(1-\ce_1-4\ce_2)/2$ & 0 \\ 
1 & $1/6$ & $1/6$ & 2/3 & 0 \\ \hline
  & $1/6$ & $1/6$ & 2/3 & 0
\end{tabular} \\ \\ \hline \hline 
\end{tabular} \\
PIRK3a: $(\ce_1, \ce_2) =  (1/4, 1/16)$ \\ 
PIRK3b: $(\ce_1, \ce_2) =  ((3-\sqrt{3})/6, (-1 +\sqrt{3})/8)$ \\
ERK3: $(\ce_1, \ce_2) = (0, 1/4)$ \\
IMEX-SSP3(4,3,3)$^*$:  $(\ce_1, \ce_2) = (0.24169426078821, (1-3 \ce_1)/4)$
\label{tab:pirk3}
\end{table}

Table~\ref{tab:pirk3} shows the tableau of the third-order PIRK method 
expressed as a four-stages IMEX-RK scheme. Note that, unlike first and 
second-order PIRK methods, the resulting IMEX-RK scheme has actually only 3 
stages because ${\tilde b_4} =0$, and the tableau could be trivially simplified.

\subsection{Fourth-order method}
\label{sect:PIRK4}
The five-stages fourth-order PIRK method can be viewed as a six-stages 
IMEX-RK scheme. For the purely explicit parts ($L1$ and $L3$ operators) we 
impose the SSP optimal five-stages fourth-order method. The remaining 
coefficients are restricted by the order conditions (see \cite{KenCar} for 
fourth-order and $s=6$) and $c_i={\tilde c}_i$ (but $b_i \neq {\tilde b}_i$ in 
general). The resulting method can be written in terms of five real 
coefficients, $\ce_i, i=1,..,5$ (do not confuse these coefficients with the 
ones appearing in previous methods), as follows:
\begin{System}
	u^{(1)} = u^n + a_{21} \, \Delta t \, L_1 (u^n, v^n), \\
	v^{(1)} = v^n + \Delta t \left[ {\tilde a}_{21} \, L_2(u^n) + \ce_1 \, L_2 (u^{(1)}) 
+ a_{21} \, L_3 (u^n, v^n) \right],
\label{e:syst4a}
\end{System}
\begin{System}
	u^{(2)} = u^n + \Delta t \left[ a_{31} \, L_1 (u^n,v^n) 
+ a_{32} \, L_1 (u^{(1)}, v^{(1)}) \right], \\
	v^{(2)} = v^n + \Delta t \left[ {\tilde a}_{31} \, L_2 (u^n) + {\tilde a}_{32} \, L_2 (u^{(1)}) 
+ \ce_2 \, L_2 (u^{(2)}) \right. \\
	\left. \hspace{2.5cm} + a_{31} \, L_3 (u^n,v^n) 
+ a_{32} \, L_3 (u^{(1)}, v^{(1)}) \right],
\label{e:syst4b}
\end{System}
\begin{System}
	u^{(3)} = u^n + \Delta t \left[ a_{41} \, L_1 (u^n, v^n) 
+ a_{42} \, L_1 (u^{(1)}, v^{(1)}) + a_{43} \, L_1 (u^{(2)}, v^{(2)}) \right], \\
	v^{(3)} = v^n + \Delta t \left[ {\tilde a}_{41} \, L_2 (u^n) + {\tilde a}_{42} \, L_2 (u^{(1)}) 
+ {\tilde a}_{43} \, L_2 (u^{(2)}) + \ce_3 \, L_2 (u^{(3)}) \right. \\
	\left. \hspace{2.5cm} + a_{41} \, L_3 (u^n,v^n) + a_{42} \, L_3 (u^{(1)}, v^{(1)}) 
+ a_{43} \, L_3 (u^{(2)}, v^{(2)}) \right],
\label{e:syst4c}
\end{System}
\begin{System}
	u^{(4)} = u^n + \Delta t \left[ a_{51} \, L_1 (u^n, v^n) 
+ a_{52} \, L_1 (u^{(1)}, v^{(1)}) + a_{53} \, L_1 (u^{(2)}, v^{(2)}) \right. \\
	\left. \hspace{2.5cm} + a_{54} \, L_1 (u^{(3)}, v^{(3)}) \right], \\
	v^{(4)} = v^n + \Delta t \left[ {\tilde a}_{51} \, L_2 (u^n) + {\tilde a}_{52} \, L_2 (u^{(1)}) 
+ \ce_4 \, L_2 (u^{(2)}) + {\tilde a}_{54} \, L_2 (u^{(3)}) \right. \\
	\hspace{2.5cm} + \ce_5 \, L_2 (u^{(4)}) \\
	\hspace{2.5cm} + a_{51} \, L_3 (u^n,v^n) + a_{52} \, L_3 (u^{(1)}, v^{(1)}) 
+ a_{53} \, L_3 (u^{(2)}, v^{(2)}) \\
	\left. \hspace{2.5cm} + a_{54} \, L_3 (u^{(3)}, v^{(3)}) \right],
\label{e:syst4d}
\end{System}
\begin{System}
	u^{n+1} = u^n + \Delta t \left[ b_{1} \, L_1 (u^n, v^n) 
+ b_{2} \, L_1 (u^{(1)}, v^{(1)}) + b_{3} \, L_1 (u^{(2)}, v^{(2)}) \right. \\
	\left. \hspace{2.5cm} + b_{4} \, L_1 (u^{(3)}, v^{(3)}) 
+ b_{5} \, L_1 (u^{(4)}, v^{(4)} ) \right], \\
	v^{n+1} = v^n + \Delta t \left[ b_{1} \, L_2 (u^n) + b_{2} \, L_2 (u^{(1)}) 
+ b_{3} \, L_2 (u^{(2)}) \right. \\
	\hspace{2.5cm} + b_{4} \, L_2 (u^{(3)}) + b_{5} \, L_2 (u^{(4)}) \\
	\hspace{2.5cm} + b_{1} \, L_3 (u^n, v^n) + b_{2} \, L_3 (u^{(1)}, v^{(1)}) 
+ b_{3} \, L_3 (u^{(2)}, v^{(2)}) \\
	\left. \hspace{2.5cm} + b_{4} \, L_3 (u^{(3)}, v^{(3)}) 
+ b_{5} \, L_3 (u^{(4)}, v^{(4)} )\right],
\label{e:syst4e}
\end{System}
where 
\begin{align*}
a_{21}&= 0.391752226571890, && && \\
a_{31}&= 0.217669096261169, &a_{32}&= 0.368410593050371, && \\
a_{41}&= 0.0826920866578107, &a_{42}&= 0.139958502191895,  &a_{43}&= 0.251891774271694, \\
a_{51}&= 0.0679662836371149, &a_{52}&= 0.115034698504631,  &a_{53}&= 0.207034898597386, \\
a_{54}&= 0.544974750228521, && && \\ 
b_1&= 0.146811876084787, &b_2&= 0.248482909444976, &b_3&= 0.104258830331981. \\
b_4&= 0.274438900901351, &b_5&=  0.226007483236906, &&
\end{align*}
are the coefficients coming from the five-stage fourth-order ERK method, and
\begin{align*}
{\tilde a}_{21} = a_{21} &- \ce_1, \\
{\tilde a}_{31} = a_{31} &+ (a_{32} - {\tilde a}_{32}) - \ce_2, \\
{\tilde a}_{32} = a_{32} &+ 0.35732150216762254 \, \ce_1 - 1.4960468621714111 \, \ce_2,  \\
{\tilde a}_{41} = a_{41} &+ (a_{42}- {\tilde a}_{42}) + (a_{43}  - {\tilde a}_{43}) - \ce_3,  \\
{\tilde a}_{42} = a_{42} &- 1.1710769982806357 \, \ce_1 + 0.5683454330255046 \, \ce_2  \\
&- 1.2113329061942606 \, \ce_3 - 1.2320330135900457 \, (a_{53} - \ce_4) \\&+ 6.103552261439627 \, \ce_5,  \\
{\tilde a}_{43} = a_{43} &- 0.37989814851159776 \, \ce_1 + 0.8235256827462162 \, (a_{53} - \ce_4)  \\
&- 4.079786814017799 \, \ce_5,  \\
{\tilde a}_{51} = a_{51} &+ (a_{52} - {\tilde a}_{52} ) + (a_{53}- \ce_4) + (a_{54} - {\tilde a}_{54}) - \ce_5,  \\
{\tilde a}_{52} = a_{52} &+ 0.1577481084030307 \, \ce_1 + 1.4709109036585493 \, \ce_3  \\
&+ 1.4960468621714111 \, (a_{53}-\ce_4) - 4.121723862609585 \, \ce_5,  \\
{\tilde a}_{54} = a_{54}&- 1.2142912127103236 \, \ce_3 + 1.432293346906654 \, \ce_5. 
\end{align*}
\begin{table}
\caption{Tableau for the explicit (left) and implicit (right) part of a general
five-stages fourth-order PIRK method, expressed as a six-stages IMEX-RK scheme.
Values of $\ce_1$ and $\ce_2$  for different schemes are also given.}
\centering
\begin{tabular}{cc} \\ \hline \hline \\
\begin{tabular}{c|cccccc}
0 & 0 & 0 & 0 & 0 & 0 & 0 \\
$c_2$ & $a_{21}$ & 0 & 0 & 0 & 0 & 0 \\ 
$c_3$ & $a_{31}$ & $a_{32}$ & 0 & 0 & 0 & 0 \\
$c_4$ & $a_{41}$ & $a_{42}$ & $a_{43}$ & 0 & 0 & 0 \\
$c_5$ & $a_{51}$ & $a_{52}$ & $a_{53}$ & $a_{54}$ & 0 & 0 \\
$1$ & $b_1$ & $b_2$ & $b_3$ & $b_4$ & $b_5$ & 0 \\ \hline
    & $b_1$ & $b_2$ & $b_3$ & $b_4$ & $b_5$ & 0
\end{tabular}
\hspace{0.5 cm}
\begin{tabular}{c|cccccc}
0 & 0 & 0 & 0 & 0 & 0 & 0 \\
$c_2$ & ${\tilde a}_{21}$ & $\ce_1$ & 0 & 0 & 0 & 0 \\ 
$c_3$ & ${\tilde a}_{31}$ & ${\tilde a}_{32}$ & $\ce_2$& 0 & 0 & 0 \\
$c_4$ & ${\tilde a}_{41}$ & ${\tilde a}_{42}$ & ${\tilde a}_{43}$ & $\ce_3$ & 0 & 0 \\
$c_5$ & ${\tilde a}_{51}$ & ${\tilde a}_{52}$ & $\ce_4$ & ${\tilde a}_{54}$ & $\ce_5$ & 0 \\
$1$ & $b_1$ & $b_2$ & $b_3$ & $b_4$ & $b_5$ & 0 \\ \hline
    & $b_1$ & $b_2$ & $b_3$ & $b_4$ & $b_5$ & 0
\end{tabular} \\ \\ \hline \hline 
\end{tabular} \\
PIRK4: $(\ce_1, \ce_2,\ce_3,\ce_4,\ce_5) = (0.13761208339219633, 0.2042433556378285,$\\
$0.0904666765339173, 0.3966145239174311, -0.00984245655482246)$ \\
ERK4: $(\ce_1, \ce_2,\ce_3,\ce_4,\ce_5) =  (0, 0, 0, a_{53}, 0)$
\label{tab:pirk4}
\end{table}

These coefficients are directly those appearing in the corresponding Butcher's 
tableau (see table~\ref{tab:pirk4}). Note that in this case, as for the 
third-order PIRK method, the resulting IMEX-RK has only 5 stages because 
$b_6=0$, and the tableau could be simplified by removing the first stage. For 
completeness we give the values of $c_i$:
\begin{align*}
 c_1&= 0, &c_2&= 0.391752226571890, &c_3&= 0.58607968931154, \\
 c_4&= 0.4745423631214, &c_5&= 0.935010630967653, &c_6&= 1.
\end{align*}
We can derive the expression for the matrix $M_4$ and its determinant, but we 
prefer to omit them in the text due to their long length. The determinant of 
the matrix results in a polynomial expression for $(\lambda \alpha_2)$ of 
degree 5 with coefficients depending on dex, trex, $a_{ij}$ and 
${\tilde a}_{ij}$; the coefficient for $(\lambda \alpha_2)^5$ only depends on 
$a_{ij}$ and ${\tilde a}_{ij}$.

We can now study numerically the behaviour of $\det(M_5)$ for 
$\omega_i=0,\pm 1$, $i=1,2$, starting from $(\lambda \alpha_2)=0$ and 
decreasing this value. Since the determinant is one scalar quantity, in 
contrast with the more complex analysis for the eigenvalues of $M_5$, the 
numerical study is much easier and very efficient. We start by setting 
$(\lambda \alpha_2)=-\epsilon$, with $\epsilon$ a positive value very close to 
0; we then minimize numerically the maximum value of $|\det(M_5)|$ for 
$\omega_i=0,\pm 1$, $i=1,2$, checking that it stays below 1; with the resulting
values for the coefficients $\ce_i$, we plot $|\det(M_5)|$ for 
$\omega_i=0,\pm 1$, $i=1,2$, and control that 
$|\det(M_5)|<1 \;\; \forall x\in[-\epsilon,0]$ for $\omega_i=0,\pm 1$, $i=1,2$;
we finally replace $\epsilon$ by a greater value and repeat the process until 
$|\det(M_5)|$ for $\omega_i=0,\pm 1$, $i=1,2$, does not stay below 1 
$\forall x\in[-\epsilon,0]$. 

The final values for the coefficients, which satisfy that 
$|\det(M_5)|<1 \;\; \forall x\in[-27,0]$ for $\omega_i=0,\pm 1$, $i=1,2$, are:
\begin{eqnarray*}
&\ce_1 = 0.13761208339219633, \; \ce_2 = 0.2042433556378285, \; \ce_3 = 0.0904666765339173, \\
&\ce_4 = 0.3966145239174311, \; \ce_5 = -0.00984245655482246.
\end{eqnarray*}
We call PIRK4 hereafter the PIRK method with these coefficients. For 
comparison, the optimal five-stage fourth-order ERK method satisfies that 
$|\det(M_5)|<1 \;\; \forall x\in[-6.75,0]$ for $\omega_i=0,\pm 1$, $i=1,2$, but
$\exists \, \omega_i=0,\pm 1$, $i=1,2$ such that $|\det(M_5)|>1$ for $x=-7$.

\subsection{Relation between the PIRK methods and the IMEX RK methods}
\label{sec:imexvspirk}

As we mentioned in Sect.~\ref{sec:pirk}, $s$-stages PIRK methods are related 
to $s+1$-stages IMEX ones. Therefore, it seems natural to compare the PIRK 
methods derived in the previous sections with IMEX RK ones of second, third and
fourth-orders existing in the literature. The relation between PIRK1 and
the IMEX$-\theta$ method has already discussed in Sect.~\ref{sec:PIRK1}.

The IMEX-SSP2(2,2,2) method presented in~\cite{ParRus} is a three-stages 
second-order scheme, that particularized to the system~(\ref{e:system}) reads:
\begin{System}
	u^{(1)} = u^n, \\
	v^{(1)} = v^n + \Delta t \, L_2(u^n),
\end{System}
\begin{System}
	u^{(2)} = u^n + \Delta t \, L_1 (u^{(1)}, v^{(1)}), \\
	v^{(2)} = v^n + \Delta t \left[ (1 - 2 \gamma) L_2(u^n) + \gamma L_2(u^{(2)})
+ L_3 (u^{(1)}, v^{(1)}) \right],
\end{System}
\begin{System}
	u^{n+1} = u^n + 0.5 \Delta t \left[ L_1 (u^{(1)}, v^{(1)}) 
+ L_1 (u^{(2)}, v^{(2)}) \right], \\
	v^{n+1} = v^n + 0.5 \Delta t \left[ L_2(u^n) + L_2(u^{(2)}) 
+ L_3 (u^{(1)}, v^{(1)}) + L_3 (u^{(2)}, v^{(2)}) \right],
\end{System}
where $\gamma = 1 - 1/\sqrt{2}$. This IMEX method is L-stable regarding the 
implicit part. In comparison with the second-order PIRK methods (PIRK2a and 
PIRK2b), there is an additional intermediate stage, which in principle implies 
additional computations. However, in practice it is only necessary to compute 
each of the differential operators twice per iteration, albeit at different 
stage levels (at $(1)$ and $(2)$ for $L_1$ and $L_3$; at $n$ and $(2)$ for 
$L_2$). This effectively transforms the method in a two-stage one (from the 
point of view of the computation of the differential operators), when applied 
to systems of separable wave-like equations. Although a smart implementation 
of this IMEX could in principle give similar computational performance per 
iteration as the PIRK2 scheme, the complexity added by the three stages of this
IMEX scheme (but only two evaluations of the differential operators) makes it 
difficult to implement efficiently in existing high performance codes. We think
that this is a good reason to make the PIRK2 methods preferable over IMEX 
ones for the case of separable wave-like equations.

The IMEX-SSP3(4,3,3) method presented in~\cite{ParRus} is a five-stages 
third-order scheme, that particularized to the system~(\ref{e:system}) reads:
\begin{System}
	u^{(1)} = u^n, \\
	v^{(1)} = v^n + \Delta t L_2(u^n),
\end{System}
\begin{System}
	u^{(2)} = u^n, \\
	v^{(2)} = v^n,
\end{System}
\begin{System}
	u^{(3)} = u^n + \Delta t \, L_1 (u^n, v^n), \\
	v^{(3)} = v^n + \Delta t \left[ (1 - \alpha) L_2(u^n) + \alpha L_2(u^{(3)})
+ L_3 (u^n, v^n) \right],
\end{System}
\begin{System}
	u^{(4)} = u^n + 0.25 \, \Delta t \left[L_1 (u^n, v^n) + L_1 (u^{(3)}, v^{(3)})
\right], \\
	v^{(4)} = v^n + \Delta t \left[\beta \, L_2(u^n) 
+ (0.5 - \alpha - \beta) L_2(u^{(3)}) + \alpha \, L_2(u^{(4)}) \right. \\
	\left. \hspace{2.5cm} + 0.25 \, L_3 (u^n, v^n) 
+ 0.25 \, L_3 (u^{(3)}, v^{(3)}) \right],
\end{System}
\begin{System}
	u^{n+1} = u^n + \frac{\Delta t}{6} \left[ L_1 (u^n, v^n) 
+ L_1 (u^{(3)}, v^{(3)}) + 4 L_1 (u^{(4)}, v^{(4)}) \right], \\ \\
	v^{n+1} = v^n + \frac{\Delta t}{6} \left[ L_2(u^n) + L_2(u^{(3)}) 
+ 4 L_2(u^{(4)}) \right. \\
	\left. \hspace{2.8cm} + L_3 (u^n, v^n) + L_3 (u^{(3)}, v^{(3)}) 
+ 4 L_3 (u^{(4)}, v^{(4)}) \right],
\end{System}
where $\alpha = 0.24169426078821$, $\beta = 0.18957643480295$. Due to the 
particular structure of the system~(\ref{e:system}), the resulting two first 
stages can be omitted, and the effective number of stages is the same as in the
case of third-order PIRK methods. The trivial value for $v^{(2)}$ results from 
a cancellation, as a consequence of the presence of opposite coefficients in 
the second stage of the numerical scheme. Actually, for this system of 
equations, the IMEX-SSP3(4,3,3) method corresponds to a third-order PIRK one, 
with $\ce_1 = \alpha = 0.24169426078821$ and 
$\ce_2 = (1-3\ce_1)/4 = 0.06872930440884$ (see also table~\ref{tab:pirk3}).
These values for the $\ce_i$ coefficients are close to one of the optimal sets 
deduced in Sect.~\ref{sect:PIRK3}, $\ce_1 = 4\ce_2 = 1/4$. As expected, both 
schemes will perform in a very similar way in the numerical examples shown in 
the next sections.

As we have shown in the previous section, the derivation of the fourth-order 
PIRK method is relatively simple, due to the fact that it is based on the 
analysis of a scalar quantity. On the contrary, the derivation of high-order 
IMEX methods when we do not focus on a particular structure of equations is 
quite complex; for example, Kennedy and Carpenter derived several schemes up to
fifth-order of convergence in~\cite{KenCar}.

\section{Application 1: system of ODEs}
\label{sect:num:ode}

Let us consider a system of ODEs of the following form:
\begin{equation}
	u_t = c\,u + d\,v, \;\;\;\; v_t = a\,u + b\,v,
\label{e:ode}
\end{equation}
where $a$, $b$, $c$ and $d$ are real constants. This system is interesting 
because it coincides with the linear part of the system of 
equations~(\ref{e:lin-system}) considered for our stability analysis, with 
$\bar{\alpha}_1=c$, $\bar{\alpha}_2=d$, $\bar{\gamma_1}=0$, $\bar{\gamma_2}=b$ 
and $\bar{\lambda}=a$. 

In the case $(b-c)^2+4\, a\,d <0$ and $b+c \le 0$, this system of equations 
has damped oscillatory solutions of the form,
\begin{equation}
	u = \frac{\sqrt{-a d}}{a} v_{0} \cos (\omega t + \phi) e^{\sigma t}, \;\;\;\;
	v = v_0 \cos(\omega t) e^{\sigma t},
\label{e:odeuv}
\end{equation}
being $v_{0}$, $\omega \equiv \frac{1}{2}\sqrt{-4\, a\, d - (b-c)^2}$, 
$\sigma \equiv \frac{b + c}{2}$ and $\tan \phi \equiv \frac{\omega}{\sigma -b}$ 
a constant set by the initial conditions, the frequency, decay rate and 
relative phase between $u$ and $v$, respectively. This system corresponds to 
(\ref{e:system}), with ${\cal L}_1 (u,v)= u+v$, ${\cal L}_2 (u)= a\,u$ and 
${\cal L}_3 (u,v) = b\,v$. Depending on the values of $\omega$, $\sigma$ and 
$\phi$, the applicability requirements of the PIRK methods given by 
definition~(\ref{def:sep-wave-like}) may impose restrictions over the possible 
values of $\Delta t$ that can be used: $\bar{\alpha}_2\, \bar{\lambda} < 0$ is 
fullfilled always; 
$|\mathrm{trex}| \leq 2 \Leftrightarrow \Delta t \leq -2/\sigma$; and 
$|\mathrm{dex}|\leq 1 \Leftrightarrow 
-2/\Delta t \leq 2\sigma + (\sigma^2 - \omega^2\cot^2\phi) \Delta t \leq 0$. We
comment here which restrictions result for the cases considered in the
numerical simulations carried out in this section. For $\sigma=0$ and 
$\phi=\pi/2$, we have no restriction for the time-step, $\Delta t < \infty$. 
For $\sigma<0$ and $\phi=\pi/2$, then $\Delta t < -2/\sigma$. For $\sigma=0$ 
and $0<\phi<\pi/2$, then $\Delta t < \frac{\sqrt{2}\tan\phi}{\omega}$.

\begin{figure}
\includegraphics[width=1.\textwidth]{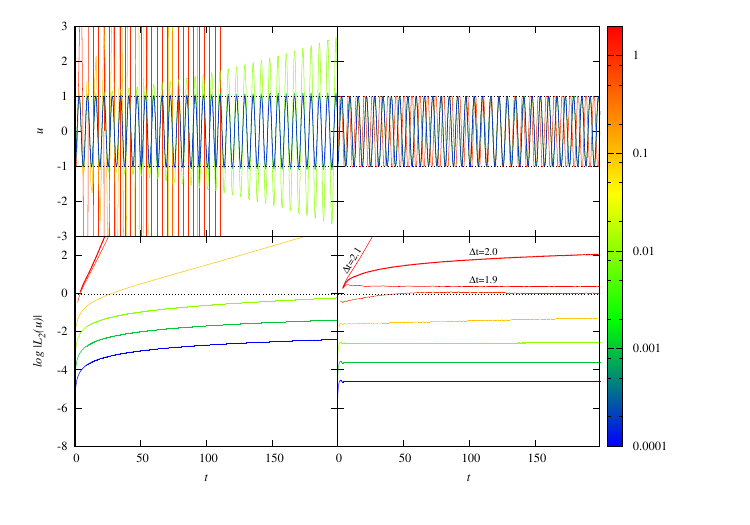} 
\caption{Numerical integration of a system of ODEs with $\sigma=\phi=0$ and 
$\phi=\pi/2$, using a first-order ERK method (left panels) and a first-order 
PIRK method (right panels). Upper panels show the time evolution of $u$, for 
$10^{-4} \leq \Delta t \leq 1$. Dotted lines are the amplitude of the 
oscillatory analytic solution. Lower panels show the time averaged $L_2$-norm 
of the difference between the numerical and analytical solutions, for 
$10^{-4} \leq \Delta t \leq 2.1$. Different lines are colored according to the 
time step, $\Delta t$, used in each simulation.}
\label{fig:ode1}
\end{figure}

For our numerical experiment we will consider the case $\omega=1$ and $a=-d$, 
without loss of generality, since it is equivalent to a rescaling of $t$ and 
$v$. The remaining coefficients depend only on the values of $\sigma$ and 
$\phi$. We have performed numerical simulations for $\sigma=0,-0.01,-0.1,-1$,
and $\phi/\pi=1/2,1/3,1/4,1/10$, which are representative of all possible 
solutions of this set of equations. As initial condition we use the analytical 
values of $u$ and $v$, given by Eq.~(\ref{e:odeuv}), at $t=0$, and $v_0 =1$.

Fig.~\ref{fig:ode1} shows the results for a representative test, comparing the 
first-order ERK method with the first-order PIRK method. To estimate the 
relative error of the method we compute the time-averaged $L_2$-norm of the 
difference between the analytic and the numerical solutions:
\begin{equation}
	L_2 (u) (t) = \frac{1}{t} 
\sqrt{\sum_{t_n<t} [u_{\mathrm{num}} (t_n) 
         - u_{\mathrm{ana}} (t_n)]^2 {\Delta t} ^2 e^{-2\sigma \, t_n}} \;\;.
\end{equation}
\begin{figure*}
\includegraphics[width=1.\textwidth]{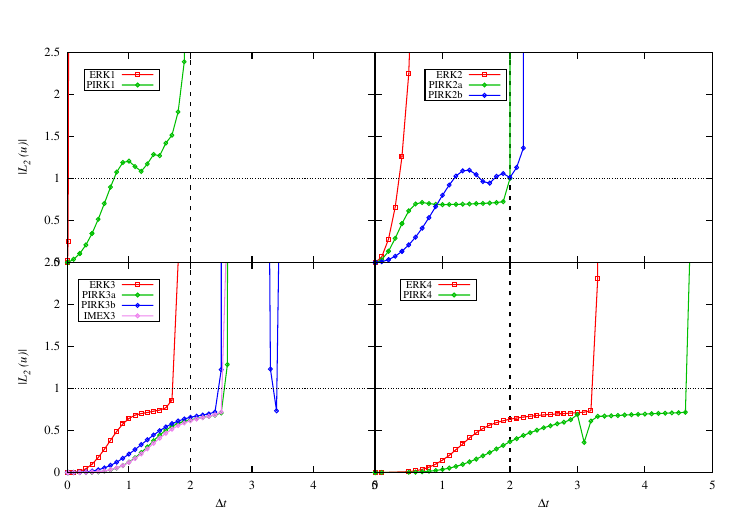} 
\includegraphics[width=1.\textwidth]{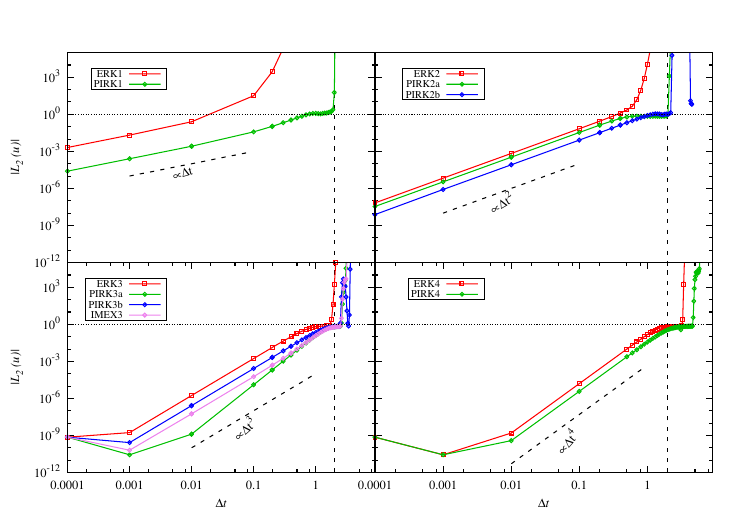} 
\caption{Numerical error integrating a system of ODEs with $\sigma=0$ and 
$\phi=\pi/2$, using first, second, third and fourth-order methods. Upper panels
show the transition between stable ($L_2 \ll 1$) and unstable ($L_2 \gg 1$) 
numerical evolutions. Lower panels, in logarithmic scale, show the behavior for
small time steps, compared to the expected scaling for each method 
(dashed-dotted lines). As a reference, vertical dashed line at $\Delta t =2$ 
corresponds to the maximum time step for the first-order PIRK method to be 
stable.}
\label{fig:ode2}
\end{figure*}
\begin{figure}
\includegraphics[width=1.\textwidth]{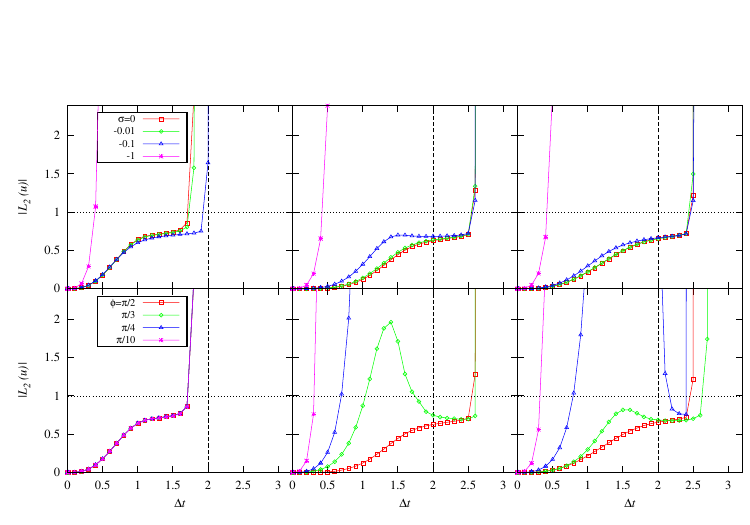} 
\caption{Behavior of the error for third-order schemes for different values of 
$\sigma$ (upper panels, $\phi=\pi/2$) and $\phi$ (lower panels, $\sigma=0$). 
From left to right, third-order ERK, PIRK3a and PIRK3b schemes.}
\label{fig:ode3}
\end{figure}

For this test the ERK method is unconditionally unstable (see left panels) and 
decreasing the time step leads to an exponentially increasing amplitude, 
provided the integration time is sufficiently long. By comparison, the 
first-order PIRK method is stable for $\Delta t < 2$, since $|u| \lesssim 1$. 
Using the PIRK method, the solution losses accuracy at late times (in this case
a phase shift), but it is still bounded (e.g. for $\Delta t = 0.1$ at 
$t=1000$), and hence the numerical method is stable.

We use the value of the time-averaged $L_2$-norm at time $t=100$ as a measure 
of the stability of a numerical method, for a particular numerical test with a 
given time step. Values $<1$ ($>1$) usually indicate stability (instability). 
In Fig.~\ref{fig:ode2} we compare the stability properties of the ERK and PIRK 
methods, for different orders of convergence, observed in our numerical 
experiments when $\sigma=0$ and $\phi=\pi/2$. In all cases, the PIRK methods 
are superior to the ERK methods, as they can achieve stable numerical 
evolutions with significantly longer time steps. For small time steps, all 
numerical methods follow the expected order of convergence. For first and 
second-order methods, ERK methods are unconditionally unstable; despite 
$L_2$-norm$<1$ for small values of $\Delta t$, longer evolutions always lead to
exponentially growing amplitudes in all studied cases. In contrast, first and 
second-order PIRKs are numerically stable in all simulations tested (up to 
$t=1000$), and only become unstable for $\Delta t$ larger than a certain 
threshold. For the third and forth-order methods, all the schemes are stable 
for small $\Delta t$, but the ERK schemes becomes unstable at lower values of 
$\Delta t$ than the PIRK ones; the third-order PIRK methods behave similar to 
the tested IMEX scheme.

A change of the value of $\sigma$, fixed $\phi=\pi/2$, introduces a damping in 
the oscillatory solution, in a timescale of $1/\sigma$. As the parameters 
approach $|\sigma \omega| \sim 1$, the system becomes stiff, and the maximum 
time-step providing stable evolutions decreases as expected. Presenting as an 
example the case of third-order methods (see upper panel of 
Fig.~\ref{fig:ode3}), and similarly for first, second and fourth-order ones, as
we approach $\sigma=-1$, both the ERK and PIRK methods behave almost 
identically. Despite of the PIRK method being {\it partially implicit}, the 
terms in Eq.~({\ref{e:ode}}) responsible for the stiffness cannot be included 
in the ${\cal L}_2$ operator, and both the ERK and PIRK methods suffer from 
this stiffness problem. 

In the case of varying $\phi$, fixed $\sigma=0$, all the ERK schemes behave in 
an identical way (see lower-right panel of Fig.~\ref{fig:ode3} for third-order 
schemes; first, second and fourth-order ones behave similarly). However, the 
PIRK methods suffer from a significant reduction of the maximum time step as 
$\phi\approx \pi/2$ (see lower-middle and right panels of Fig.~\ref{fig:ode3} 
for third-order schemes; first, second and fourth-order ones behave similarly).
This is the only case in which the ERK methods are superior to the PIRK 
methods. Therefore, the class of systems for which the PIRK methods are a good 
alternative to the classical ERK methods are wave-like equations, in which the 
condition $\phi\approx\pi/2$ is fulfilled.

\section{Application 2: Scalar wave equation}
\label{sect:num:sph}

Our second test case is the time evolution of a scalar wave equation in 
spherical coordinates. Our motivation for choosing this test is the 
difficulties that the numerical relativity community has traditionally 
encounter evolving Einstein equations in these particular set of coordinates. 
The case of Einstein equations involves the evolution of non-linear wave-like 
equations for tensorial quantities. To elude the difficulties of evolving such 
systems of equations and have a clean test of our numerical methods with 
analytic exact solutions, we have chosen here to study the evolution of an 
scalar wave equation. Therefore, we have concentrated most of our effort in 
this particular test, for which the most detailed analysis is given here. The 
wave equation in Cartesian coordinates is also studied, at the end of this 
section, albeit less thoughtfully, to check that the stability properties of 
the system of equations do not depend on the system of coordinates (as it 
should be from the analysis carried out in Sect.~\ref{sect:met-stab}).

A wave equation for a scalar $h$ can be written as:
\begin{equation}
	\partial_{tt} h = \triangle h,
\label{e:wave}
\end{equation}
where $\triangle$ is the Laplacian operator. Eq.~(\ref{e:wave}) can be 
rewritten as a first-order system in time, with the addition of an extra 
auxiliary variable, $A$, as follows:
\begin{eqnarray}
	\partial_t h &=& A, \nonumber \\
	\partial_t A &=& \triangle h.
\label{e:wave2}
\end{eqnarray}
In this case, according to system~(\ref{e:system}), the variables can be 
identified as $(u, v) = (h, A)$, and the operators as ${\cal L}_1(h, A) = A$, 
${\cal L}_2(h) = \triangle h$ and ${\cal L}_3(h, A) = 0$. Therefore,
$\bar{\alpha}_1 = \bar{\gamma}_1 = \bar{\gamma}_2 = 0$ and $\bar{\alpha}_2 = 1$
in system~(\ref{e:lin-system}). The eigenvalues of the linearized explicit part
are $\omega_1 = \omega_2 = 1$ and hence dex $=1$ and trex $= 2$. 
$\bar{\lambda} \in \mathbb{R}^-$ and its value depends on the particular 
solution for $h$ and the discretization of the operator $\triangle$. Spherical 
coordinates are used, being $\displaystyle \left(\frac{\partial}{\partial r}, 
\frac{1}{r}\frac{\partial}{\partial \theta}, 
\frac{1}{r \sin\theta}\frac{\partial}{\partial \varphi}\right)$ the 
corresponding orthonormal basis. 

Eq.~(\ref{e:wave}) has solutions of the form
\begin{equation}
	h (r, \theta,\varphi, t) \sim  j_l (k r) Y_{lm} (\theta, \varphi) \cos{k t},
\label{eq:wavesol}
\end{equation}
being $j_l$ the spherical Bessel function of first kind of order $l$ and 
$Y_{lm}$ the spherical harmonics. The value of $k$, a positive real constant, 
is determined by imposing boundary conditions. We search for solutions inside a
sphere of radius unity imposing $h(r = 1, \theta, \varphi, t) = 0$. For fixed 
$l$ and $m$, it is possible to compute the eigenmode frequencies, $k_{nl}$, as 
the zeros of the spherical Bessel function of order $l$, being $n=1$ the first 
zero and so on.

We have performed 1D-spherical, 2D-axisymmetric and 3D simulations of the 
system using as initial data solutions with $n=1$ at $t=0$. We use values of 
$(l,m) = (0,0), (2,0), (2,2)$, for the 1D, 2D and 3D case, respectively. In 
this way, the initial data are regular and fulfill the symmetries in each case.
Furthermore, the data are non trivial for each symmetry, in the sense that 
there are not trivial cancellations for 2D and 3D simulations. We consider 
symmetry with respect to the equatorial plane, $\theta = \pi/2$. 

We use a finite difference scheme to solve the system using an equally-spaced 
grid with $n_r$, $n_\theta$ and $n_\varphi$ grid points in the $r$, $\theta$ 
and $\varphi$ directions, respectively. We use derivatives of cell-centered 
Lagrange interpolation polynomials~\cite{polLa} to compute the first and second
spatial derivatives appearing in the Laplacian operator, achieving fourth, 
sixth and eighth discretization order. Boundary conditions are imposed by using
a number of ghost cells consistent with the discretization stencil. The 
analytical solution is imposed as boundary condition at $r=1$. Symmetry 
conditions are used at all other boundaries. The maximum time step is 
determined by the CFL condition for the speed of the wave, which is 1. The time
step in the simulations is a smaller fraction of the maximum time step, the CFL
factor, in the interval $[0,1]$.
 
We can estimate the absolute error of the numerical evolution by comparing the 
numerical solution with the analytical one given by Eq.~(\ref{eq:wavesol}). As 
a measure of the global error during the numerical evolution, we compute the 
$L_2$-norm of the difference between the numerical and the analytical solutions
at a given time,
\begin{equation}
	L_2(h)(t) = \frac{1}{n_r n_\theta n_\varphi} \sqrt{ 
\sum_{r, \theta, \varphi} \left[ 
h_{\rm num}(r, \theta, \varphi, t) - h_{\rm ana}(r, \theta, \varphi, t) 
\right]^2 \, (k \, r)^2}.
\end{equation}

\subsection{Stability}

We have studied numerically the stability of the first, second and third-order 
PIRK methods. This study involves the numerical computation of a wide parameter
space, including the coefficients $\ce_i$ of the PIRK methods and the CFL 
factor. Up to 10000 simulations have to be performed to cover this parameter 
space, so we have decided to use a single numerical setup as reference for the 
stability study. We use $(n,l,m)=(1,2,0)$ for the initial data in 
2D-axisymmetry with equatorial symmetry. In this case, $k_{12} \approx 5.763$. 
We use $(n_r, n_\theta) = (100, 32)$ grid points and a fourth-order spatial 
discretization scheme. 

With the numerical setup fixed, we can estimate the value of 
$x := - \lambda \alpha_2 = - \bar{\lambda} (\Delta t)^2$, which is the relevant
quantity for the stability of the system. For solutions of the 
form~(\ref{eq:wavesol}), Eq.~(\ref{e:wave2}) can be written as
\begin{eqnarray}
	\partial_t h &=& A, \nonumber \\
	\partial_t A &=& -k^2 h.
\end{eqnarray}
Therefore, $\bar{\lambda} = -k^2$. The minimum value of $k$ corresponds to the 
fundamental mode $(n,l,m)=(1,0,0)$, i.e. $k_{\rm min} = k_{10}= \pi$. This sets
the lower limit for $x$,
\begin{equation}
	x_{\min} = \pi^2 (\Delta t)^2.
\end{equation}
Note that in the limit of infinite resolution, the CFL restriction results in 
$\Delta t \to 0$ and $x_{\min} \to 0$. We show in the next subsections that 
this is indeed the case for the typical resolutions used in practical 
applications.

The maximum value of $k$ corresponds to solutions with typical spatial 
variations of the order of the smallest grid cell size, $\Delta l_{\rm min}$. 
For these solutions,
\begin{equation}
	k_{\rm max} \approx \frac{2 \pi}{\Delta l_{\rm min}},
\end{equation}
and hence the upper limit for $x$ is
\begin{equation}
	x_{\rm max} = k_{\max}^2 (\Delta t)^2  \approx 
\left( \frac{2 \pi \Delta t_{\rm max}}{\Delta l_{\rm min}} \right)^2 
({\rm CFL \,\, factor})^2 = 4 \pi^2 ({\rm CFL \,\, factor})^2,
\end{equation}
being the ${\rm CFL \,\, factor} = \Delta t / \Delta t_{\rm max}$ and 
$\Delta t_{\rm max} = \Delta l_{\rm min}$ the maximum time step allowed by the 
CFL condition, which coincides with the smallest grid cell size. Note that 
$x_{\rm max}$ does not depend on the resolution. As a consequence, the 
stability properties of the system do not depend on the resolution, but only on
the CFL factor. For convenience we define 
$\bar x := x / ({\rm CFL \,factor})^2$; therefore, 
$\bar{x}_{\rm min} = \pi^2 (\Delta t_{\rm max})^2$ and 
$\bar{x}_{\rm max} = 4 \pi^2$.
By keeping the numerical setup fixed, we expect that the limits of $\bar{x}$ 
stay constant in all our simulations. Effects of the dimensionality and 
resolution are discussed in Sect.~\ref{sect:con}.

For each order of the PIRK methods, we compare stability behavior of the 
numerical simulations with the stability criterion of Sect.~\ref{sect:met-stab}
for the determinant of the corresponding matrix of a given method. We compare 
this stability criterion with the general one for the eigenvalues of the 
system. 

\subsubsection{First-order method}
\label{sec:num1}

The expression for $\det(M_1)$ particularized to the system~(\ref{e:wave2}) 
with $x>0$, is 
\begin{equation}
	\det M_1 = 1 + (1 - \ce_1) x,
\end{equation}
and the eigenvalues of $M_1$ are
\begin{equation}
	e_\pm = \frac{1}{2} \left ( 2 - \ce_1 x \pm \sqrt{-x (4 - \ce_1^2 x)} \right).
\label{e:eigen1}
\end{equation}
The stability criterion, $|e_\pm| \leq 1$, for $x \ne 0$, results in 
\begin{eqnarray}
	1 \le &\ce_1& \le \frac{1}{2} + \frac{2}{x}, \label{e:numbound1a}\\ 
	x &\le& 4. \label{e:numbound1b}
\end{eqnarray}
The condition $|\det{M_1}| \leq 1$, used in Sect.~\ref{sect:met-stab}, results 
in 
\begin{equation}
	1 \le \ce_1 \le 1 + \frac{2}{x},
        \label{e:numbounddet1}
\end{equation}
which is a necessary, but not sufficient, condition for stability. 
Conditions~(\ref{e:numbound1a})--(\ref{e:numbounddet1}) are more restrictive 
for larger values of $x$, so the value $\bar{x}_{\rm max}$ determines the 
stability properties of the system.

\begin{figure}
\begin{center}
	\includegraphics[width=0.491\textwidth]{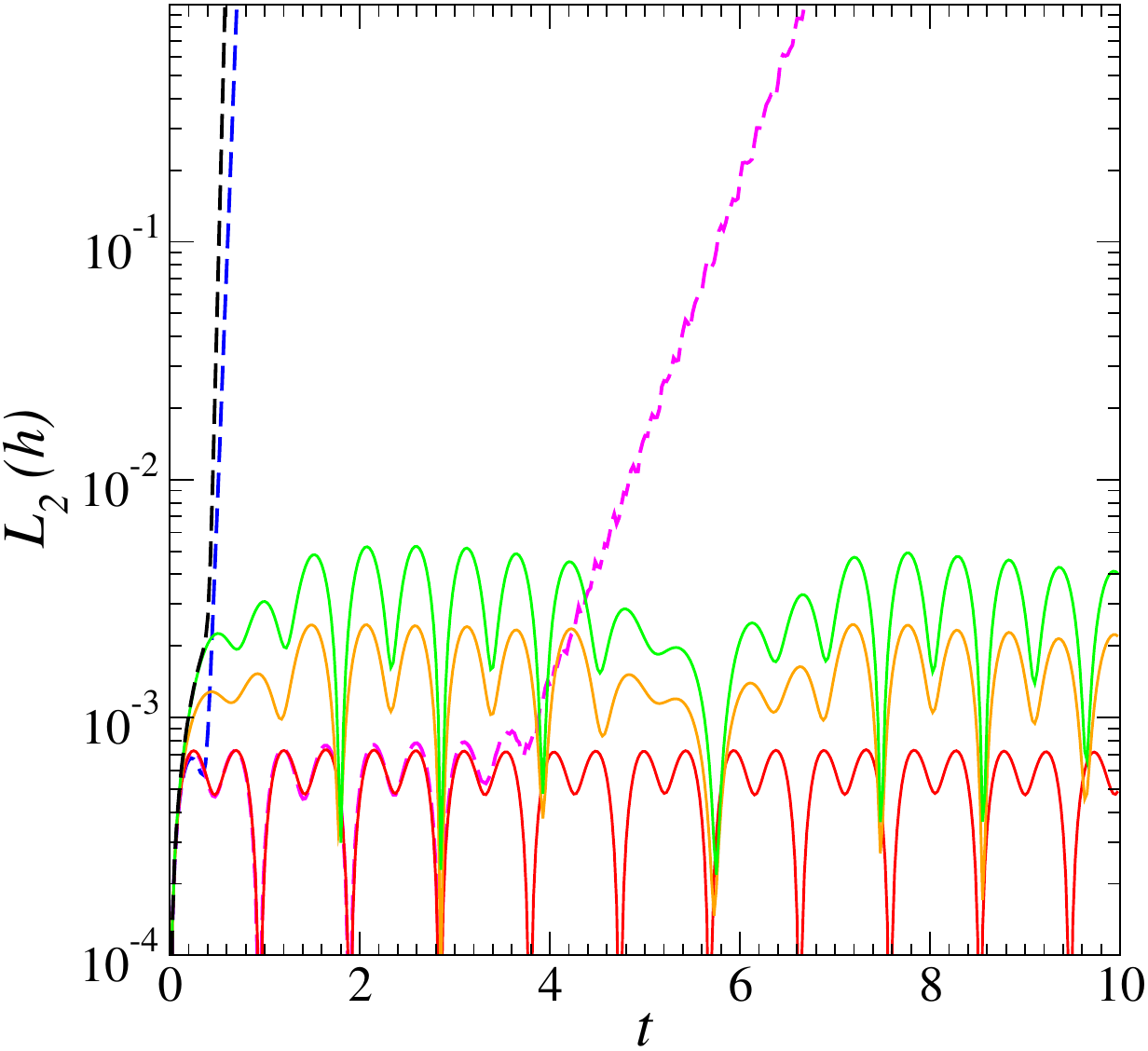} 
	\hspace{0.2cm}
	\includegraphics[width=0.45\textwidth]{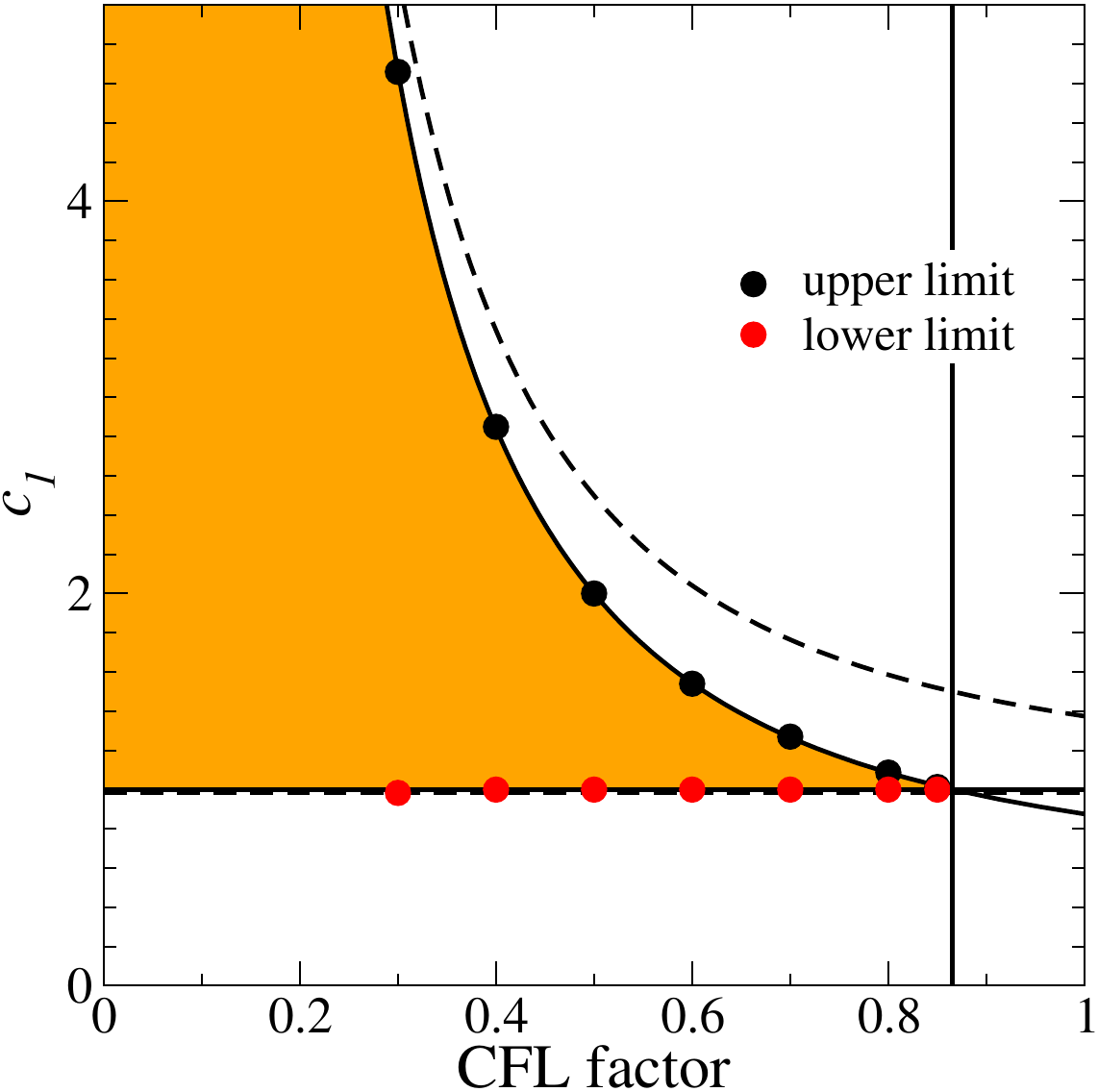} 
	\caption{Stability of the first-order PIRK method. {\it Left panel}: time 
evolution of the $L_2$-norm for simulations with CFL factor $0.5$ and $\ce_1$ 
values of $0.9$ (blue), $0.99$ (magenta), $1$ (red), $1.5$ (orange), $2$ 
(green) and $2.05$ (black). Solid and dashed lines represent numerically stable
and unstable simulations, respectively. {\it Right panel:} stability region 
depending on the values for $\ce_1$ and the CFL factor. Red and black circles 
are the maximum and minimum values of $\ce_1$, respectively, for which a 
numerically stable evolution is found for each CFL factor. Solid lines are the 
boundaries of the stability region (orange area) given by 
Eqs.~(\ref{e:numbound1a}) and (\ref{e:numbound1b}), for the fitted value of 
$\bar{x} = 5.340 $. The boundary of the region $|\det(M_1)| \leq 1$, which 
partially coincides with the stability boundaries, is also plotted (dashed 
lines).}
\label{fig:1}
\end{center}
\end{figure}

It is therefore relevant to study the stability properties of the numerical 
solution depending on the coefficient $\ce_1$ and the time step $\Delta t$. We 
have performed series of simulations for several CFL factors ($0.3$, $0.4$, 
$0.5$, $0.6$, $0.7$, $0.8$, $0.85$ and $0.9$), varying the value of $\ce_1$ 
from $0$ to $5$ in steps of $0.001$ ($4\times 10^4$ simulations in total). We 
have evolved each simulation up to time $t=54$, i.e. about $50$ oscillations. 
The left panel of Fig.~\ref{fig:1} shows the evolution of the $L_2$-norm for a 
selection of simulations. The behavior of this quantity allows us to 
distinguish between numerically stable and unstable evolutions in a similar 
fashion as we did for the ODE case of Sect.~\ref{sect:num:ode}. Numerically 
stable evolutions (solid lines) show an oscillatory behavior of the $L_2$-norm 
with values smaller than one, while numerically unstable evolutions (dashed 
lines) show an exponential grow of the $L_2$-norm which becomes much larger 
than $1$. For the purpose of this test, we consider a numerical simulation to 
be numerically stable at a point of the parameter space, given by the CFL 
factor and the coefficients of the PIRK method, if $L_2$-norm$<1$ at a time 
corresponding to $50$ oscillations, i.e. $t=100\pi/\omega$. 

For each CFL factor, we find a maximum and minimum values of $\ce_1$ such that 
all simulations within this two values are numerically stable, while lower or 
higher values lead to numerically unstable evolutions. No numerically stable 
values of $\ce_1$ were found for a CFL $=0.9$. The right panel of 
Fig.~\ref{fig:1} shows the numerically determined stability limits (red and 
black circles). We can compare these limits with the predictions of 
Eqs.~(\ref{e:numbound1a}) and (\ref{e:numbound1b}). For all CFL factors the 
lower limit is $\ce_1=1$ and coincides with the prediction of 
Eq.~(\ref{e:numbound1a}). Following Eq.~(\ref{e:numbound1a}), we fit the upper 
bound by a curve of the form $p_1 + p_2  (\Delta t_{\rm max} / \Delta t)^2$, 
with $p_1=0.501$, $p_2=0.3746$ and $\Delta t_{\rm max}$ the maximum time step 
predicted by the CFL condition. We confirm the $1/(\Delta t)^2$ behavior of the
upper limit and a value of $p_1$ compatible with the predicted $1/2$ in 
Eq.~(\ref{e:numbound1a}). From the fitted value of $p_2$ we estimate 
$\bar{x} = 5.340$, which is of the order of magnitude of 
$\bar{x}_{\rm max} \approx \pi^2$. Using this estimation for $\bar{x}$, it is 
possible to compute the maximum CFL factor using the condition 
(\ref{e:numbound1b}). It results to be $0.8656$, consistent with not finding 
any stable simulation for CFL $=0.9$.

If we consider the stability criterion used in Sect.~\ref{sec:PIRK1} 
(dashed-line in Fig.~\ref{fig:1}) for the determinant of $M_1$, we observe that
this criterion is overestimating the stability region. However, the estimated 
optimal value, $\ce_1=1$, lays inside the stability region and is indeed the 
value such that the maximum CFL factor is achievable.

The first-order ERK method can be recovered by setting $\ce_1=0$
\footnote{Alternatively, identical result can be obtained by setting 
$\bar \alpha_1 = \bar \gamma_1 = \bar \lambda = 0$, $\bar \alpha_2 = 1$ 
and $\bar \gamma_1 \ne 0$ for all of the PIRK methods presented in this work.}.
Since none of our numerical simulations with $\ce_1=0$ and CFL factors between 
$0.3$ and $0.95$ show stable evolutions (see right panel of Fig.~\ref{fig:1}), 
we have performed simulations decreasing the CFL factor to try to find the 
stability limit. We were not able to find stable evolutions for CFL factors as 
low as $0.001$. The reason for this behavior is that the eigenvalues of the 
first-order ERK method, given by Eq.~(\ref{e:eigen1}) with $\ce_1=0$, read 
\begin{equation}
	e_{\pm} = 1 \pm \sqrt{-x},
\end{equation}
and the stability criterion, $|e_\pm| \le 1$, is only fulfilled for $x=0$, i.e.
$\Delta t = 0$. If we decrease the CFL factor, i.e. approach $|e_\pm|=1$, the 
instability appears at later time in the simulation (e.g., for CFL $= 0.001$ 
the instability appears at $t \sim 7$). The consequence is that first-order ERK
method can be used for finite-time evolutions provided a sufficiently small CFL
factor is used. However, the time step restriction of the first-order ERK 
method is significantly larger (several orders of magnitude) than the one of 
the first-order PIRK scheme.

\subsubsection{Second-order method}
\label{sec:num2}

The expression for $\det(M_2)$ particularized to the system~(\ref{e:wave2}) 
with $x>0$, is
\begin{equation}
	\det(M_2) = \frac{1}{4} \left[ 4 -x (2\ce_2 - \ce_1 - 2\ce_1 \ce_2) \right],
\end{equation}
and the eigenvalues of $M_2$ are
\begin{eqnarray}
	e_\pm &=& 1 - \frac{x}{2} + \frac{\ce_1}{8} (1 - 2 \ce_2) \, x^2  \nonumber \\
	&\pm& \frac{1}{8}\sqrt{-x \left [64 - 16 (1 + 2 \ce_1 - 2 \ce_2) x 
+ 8 \ce_1 (1 - 2 \ce_2) x^2 - \ce_1^2 (1 - 2 \ce_2)^2 x^3\right ]}.
\end{eqnarray}
The stability condition, $|e_\pm| \leq 1$, leads to 
\begin{eqnarray}
	&& \frac{4}{x} \left(1 - \frac{4}{x}\right) \le \ce_1 (1-2\ce_2) \le \frac{4}{x},
\label{e:numbound2a}\\
	&& \ce_1 - \ce_2 \le 2/x, 
\label{e:numbound2b}\\
	&& \ce_2 (2 \ce_1 - 1) \le \frac{2}{x} \left(-1 + \frac{4}{x} \right),
\label{e:numbound2c}\\
	&& 0 \le \ce_1 + 2 \ce_2 (\ce_1 - 1) .
\label{e:numbound2d}
\end{eqnarray}
The condition $|\det(M_2)| \leq 1$ is equivalent to
\begin{equation}
	0 \leq \ce_1 + 2 \ce_2 (\ce_1 - 1) \leq \frac{8}{x^2},
        \label{e:numbounddet2}
\end{equation}
which coincides partially with the boundaries of the stability region.
The conditions~(\ref{e:numbound2a})-(\ref{e:numbound2d}) and 
(\ref{e:numbounddet2}) are more restrictive for larger values of $x$, 
therefore the value $\bar{x}_{\rm max}$ determines the stability properties
of the system. 

\begin{figure}
\begin{center}
	\includegraphics[width=0.49\textwidth]{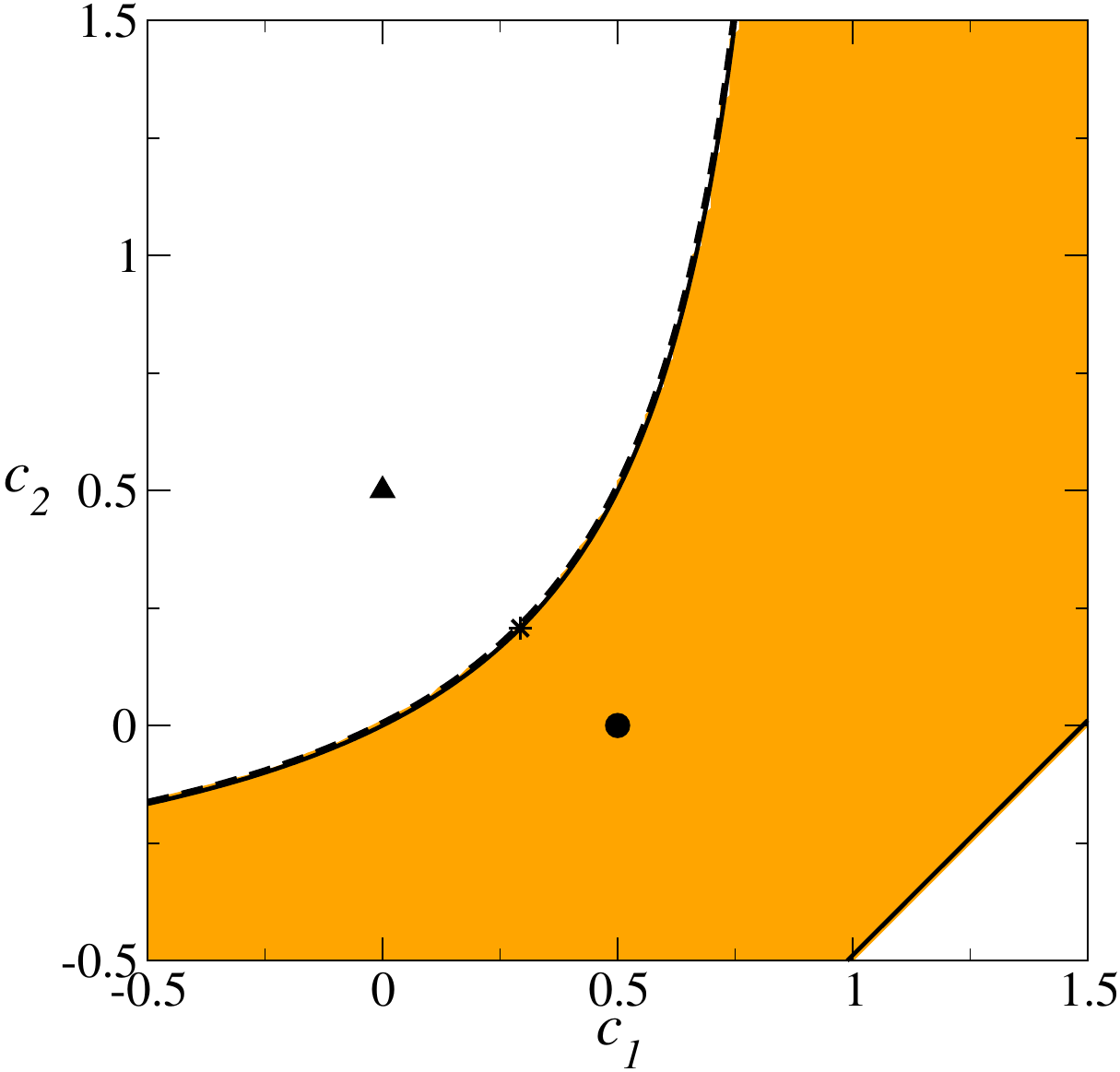}
	\hspace{0.05cm}
	\includegraphics[width=0.49\textwidth]{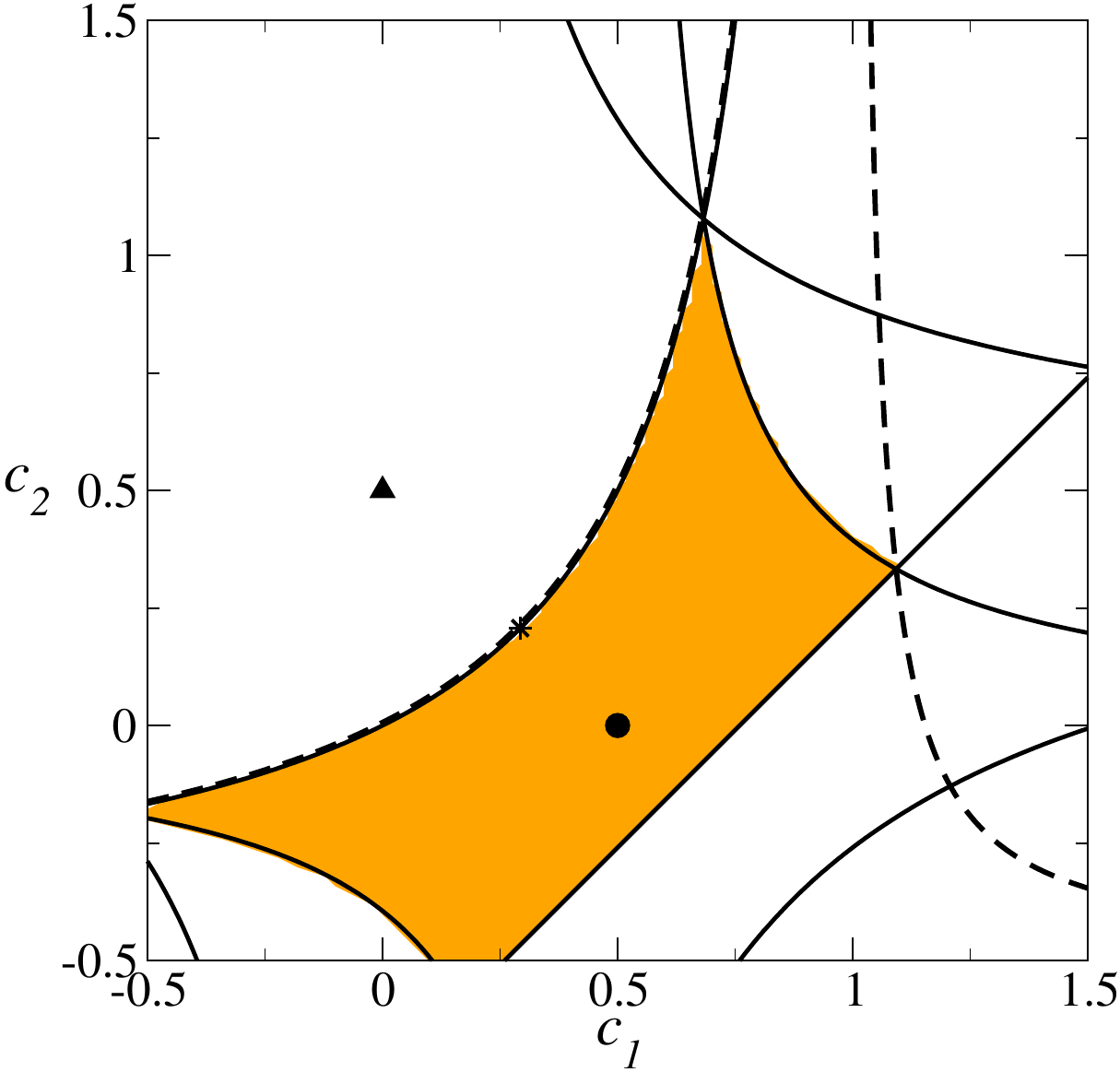}
	\\ \vspace{0.14cm}
	\includegraphics[width=0.49\textwidth]{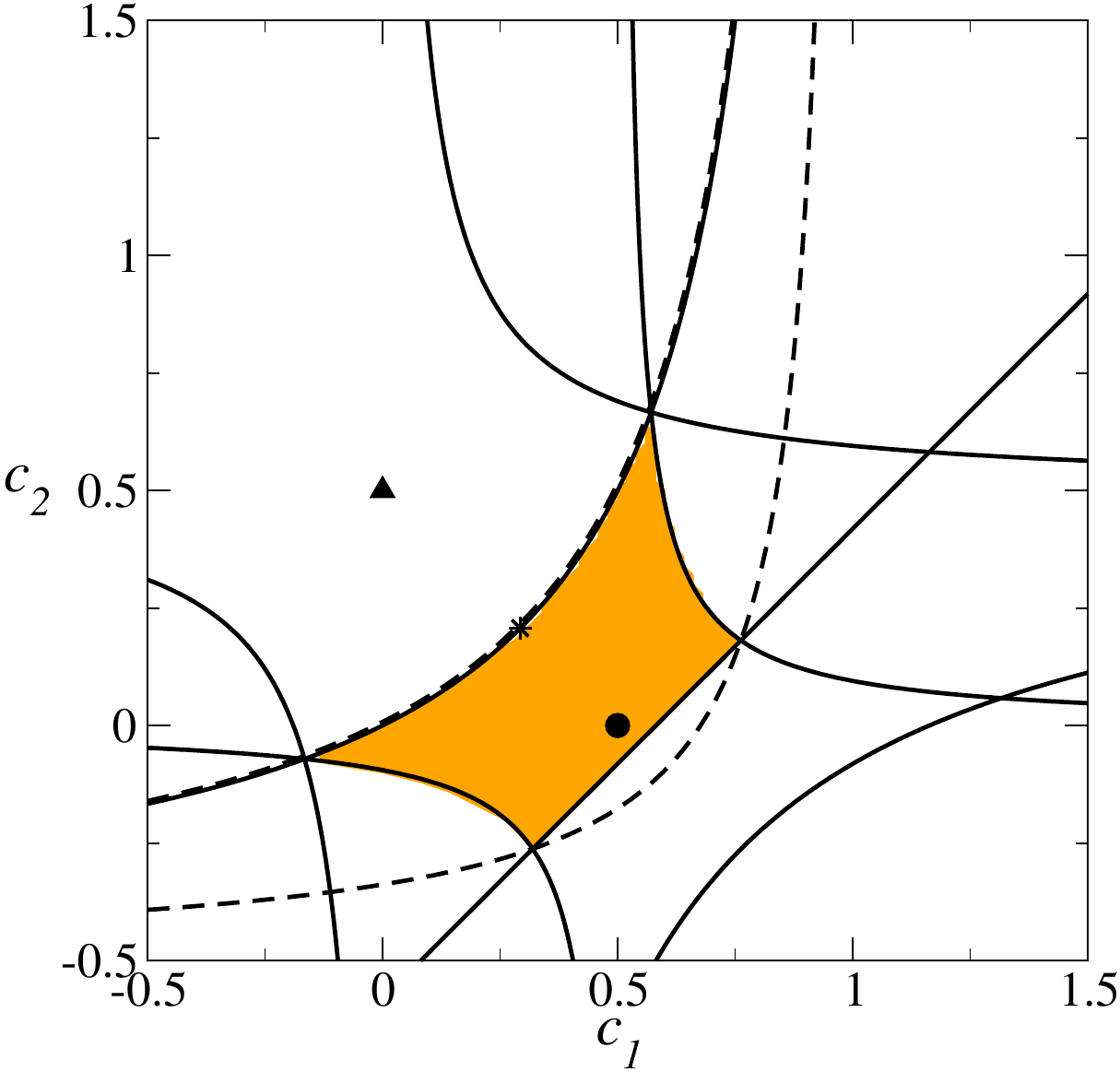}
	 \hspace{0.05cm}
	\includegraphics[width=0.49\textwidth]{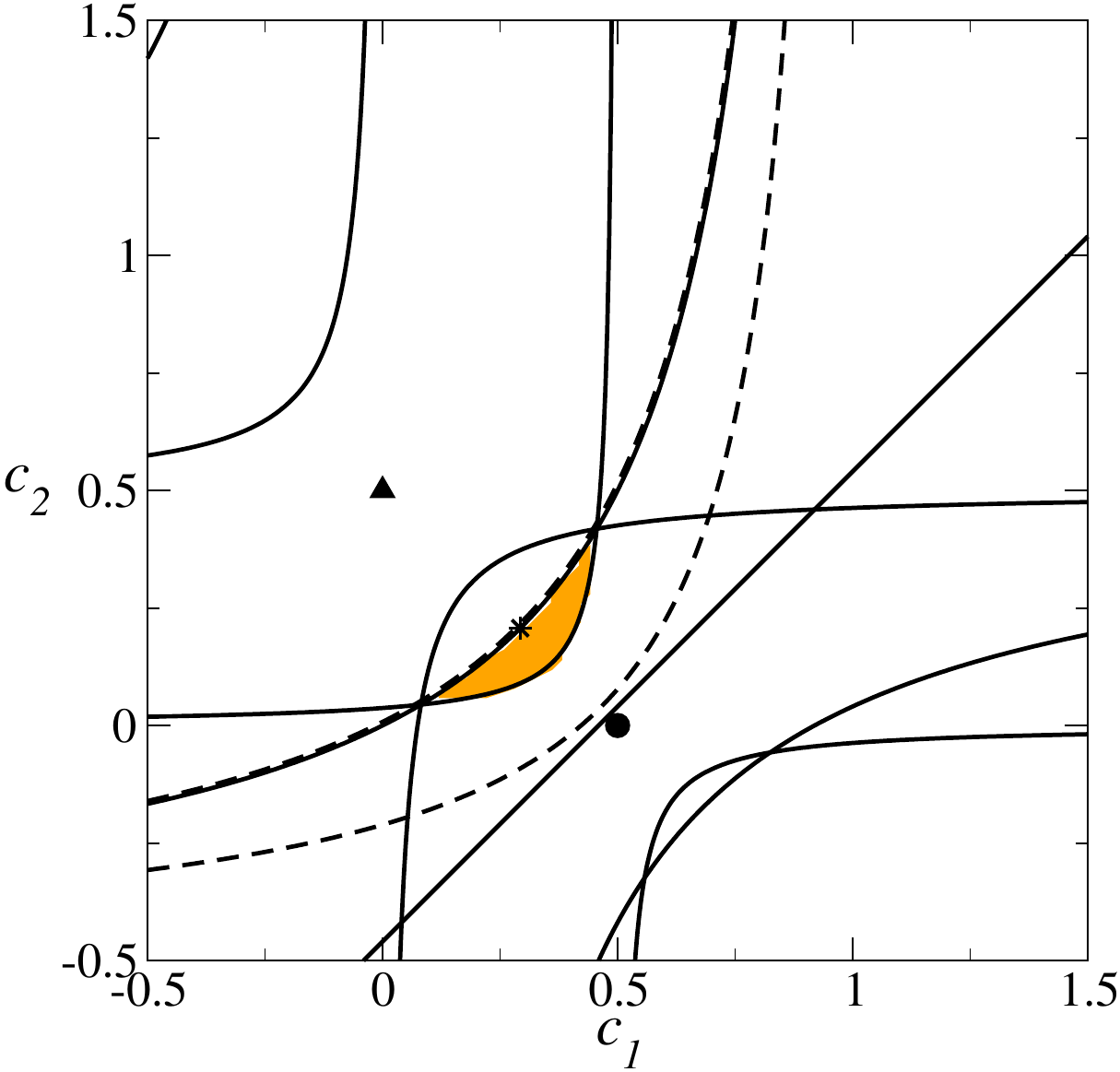}
	\caption{Dependence of the numerically determined stability region (orange 
	area) on the $(\ce_1, \ce_2)$ coefficients for the evolution of $h$ using a 
	second-order PIRK method. The boundaries (solid lines) of the stability 
	region according to Eqs.~(\ref{e:numbound2a})-(\ref{e:numbound2d}) for 
	$\bar{x} = 5.373$, and the boundaries (dashed lines) of the region 
	$|\det(M_2)| \leq 1$, which partially coincide with the stability boundaries, 
	are plotted. The values for the optimal second-order PIRK method with 
	$(\ce_1, \ce_2) = (1/2, 0)$ (PIRK2a, black circle) and 
	$(\ce_1, \ce_2) = 1/2 (2 - \sqrt{2}, \sqrt{2} - 1)$ (PIRK2b, star symbol), 
	and the second-order ERK scheme (black triangle) are also plotted.}
\label{fig:2}
\end{center}
\end{figure}

We have studied the numerical stability of the second-order PIRK method by 
performing simulations using $\ce_1 \in [-0.5,1.5]$ and 
$\ce_2 \in [-0.5, 1.5]$, varying the coefficients in steps of $0.02$. We have 
used several CFL values ($0.5$, $0.7$, $0.8$ and $0.9$). Fig.~\ref{fig:2} shows
the stability region on the $(\ce_1,\ce_2)$ plane. Points of this plane leading
to numerically stable simulations are plotted in orange color in 
Fig.~\ref{fig:2}, while numerically unstable simulations ($L_2$-norm$>1$ at 
$t=100\pi/\omega$) are plotted in white. We use the same convention for the 
third-order method tests (Fig.~\ref{fig:3}).

According to Eqs.~(\ref{e:numbound2a})-(\ref{e:numbound2d}), the boundaries of 
the stability region depend only on the parameter $\bar{x}$. In order to 
estimate $\bar{x}$, we perform a $\chi^2$-minimization of the difference 
between the numerically determined boundaries for all CFL factors and the 
theoretically predicted values. We obtain $\bar{x} = 5.373$ with high 
significance ($\chi^2 = 353.7$ for $385$ degrees of freedom). Note that, as 
expected, the value of $\bar x$ is very close to the one obtained in the 
first-order PIRK method. The boundaries for the fitted value of $\bar x$ 
(solid-lines in Fig.~\ref{fig:2}) agree very closely with the numerically 
computed ones. As in the first-order PIRK method, the criterion 
$|\det{M_2}| \leq 1$ overestimates the stability region (dashed-lines in 
Fig.~\ref{fig:2}).

We can compare the numerical results for the stability region with the optimal 
values computed in Sect.~\ref{sec:PIRK2}: $(\ce_1,\ce_2) = (1/2,0)$ for 
$|x| \ll 1$ (PIRK2a, black circle in Fig.~\ref{fig:2}) and 
$(\ce_1,\ce_2) = \frac{1}{2} (2-\sqrt{2}, \sqrt{2}-1)$ for $|x| \gg 1$ (PIRK2b,
star symbol in Fig.~\ref{fig:2}). Both values allow for stable numerical 
evolutions with CFL factors close to unity. According to the fit, 
$x = 5.373 \, ({\rm CFL\,factor})^2$, which is larger than unity for CFL 
factors larger than $0.434$. Indeed, the PIRK2b is better suited for higher CFL 
values, while the PIRK2a becomes unstable. Therefore, we recommend the optimal 
values $(\ce_1, \ce_2)=\frac{1}{2}(2-\sqrt{2},\sqrt{2}-1)$ (PIRK2b) as they 
seem to provide stable evolutions with the largest possible CFL factors.

The second-order ERK scheme corresponds to the case $(\ce_1,\ce_2)=(0,1/2)$ and
is unconditionally unstable. In this case, the eigenvalues are
\begin{equation}
	e_\pm = 1 -\frac{x}{2} \pm \sqrt{-x},
\end{equation}
and the stability criterion, $|e_\pm|\le 1$, is only fulfilled for 
$\Delta t = 0$. Therefore, the observed numerical behavior of the second-order 
ERK scheme is very similar to the first-order ERK one.

\subsubsection{Third-order method}
\label{sec:num3}

The expression for $\det(M_3)$ particularized to the system~(\ref{e:wave2}) 
with $x>0$ is 
\begin{equation}
  \det(M_3) = 1 + \frac{x^2}{12} (\ce_1 - 4 \ce_2) + 
\frac{x^3}{72} [-1 + 3 (1 - 2 \ce_1) (\ce_1 + 4 \ce_2)],
\end{equation}
and the eigenvalues of $M_3$ are
\begin{eqnarray}
	e_\pm &=& 1 - \frac{x}{2} + \frac{x^2}{24} (1 + \ce_1 - 4 \ce_2) 
- \frac{\sqrt{x}}{24} \left[ \, 192 \, (x - 3) \right. \nonumber\\
	&&\left. - 16 x^2 \left(3 \ce_1 (1 - c1 - 4 \ce_2) + 1 \right) 
+ x^3 (1 + \ce_1 - 4 \ce_2)^2 \right ]^{1/2}.
\end{eqnarray}
The stability criterion, $|e_\pm| \leq 1$, for $x\ne 0$ results in 
\begin{eqnarray}
	\frac{12}{x} \left(1-\frac{4}{x}\right) &\le& 1 + \ce_1 - 4\ce_2 \le \frac{12}{x},
\label{e:numbound3a}\\
  1 + 3 (2 \ce_1 - 1) (\ce_1 + 4\ce_2) &\le& \frac{6}{x} \left(\frac{12}{x}-1\right),
\label{e:numbound3b}\\
  1 + 3 (2 \ce_1 - 1) (\ce_1 + 4\ce_2) &\le& \frac{6}{x} \left[
\frac{12}{x} \left(\frac{4}{x} - 1\right) - 1 + 2 (1 + \ce_1 - 4 \ce_2) \right],
\label{e:numbound3c}\\
  \frac{6}{x} (\ce_1 - 4 \ce_2) &\le& 1 + 3 (2 \ce_1 - 1) (\ce_1 + 4 \ce_2).
\label{e:numbound3d}
\end{eqnarray}
The condition $|\det(M_3)| \leq 1$ is equivalent to
\begin{eqnarray}
  \frac{6}{x} (\ce_1 - 4 \ce_2) &\le& 1 + 3 (2 \ce_1 - 1) (\ce_1 + 4 \ce_2),
  \label{e:numbounddet3a} \\
  -144 &\le& 6 x^2 (\ce_1 - 4 \ce_2) + x^3 [-1 + 3 (1 - 2 \ce_1) (\ce_1 + 4 \ce_2)],
  \label{e:numbounddet3b}
\end{eqnarray}
which coincides partially with the boundaries of the stability region. The 
conditions~(\ref{e:numbound3a})-(\ref{e:numbound3c}) and 
(\ref{e:numbounddet3b}) are more restrictive for larger values of $x$, 
therefore the value $\bar{x}_{\rm max}$ determines the stability properties of 
the system at these boundaries. However the condition (\ref{e:numbound3d}) 
(which is equivalent to (\ref{e:numbounddet3a})), can be more restrictive for 
small values of $x$ depending on the value of $\ce_1$ and $\ce_2$. For this 
case, both $\bar{x}_{\rm max}$ and $\bar{x}_{\rm min}$ are relevant for the 
stability analysis.

\begin{figure}
\begin{center}
	\includegraphics[width=0.49\textwidth]{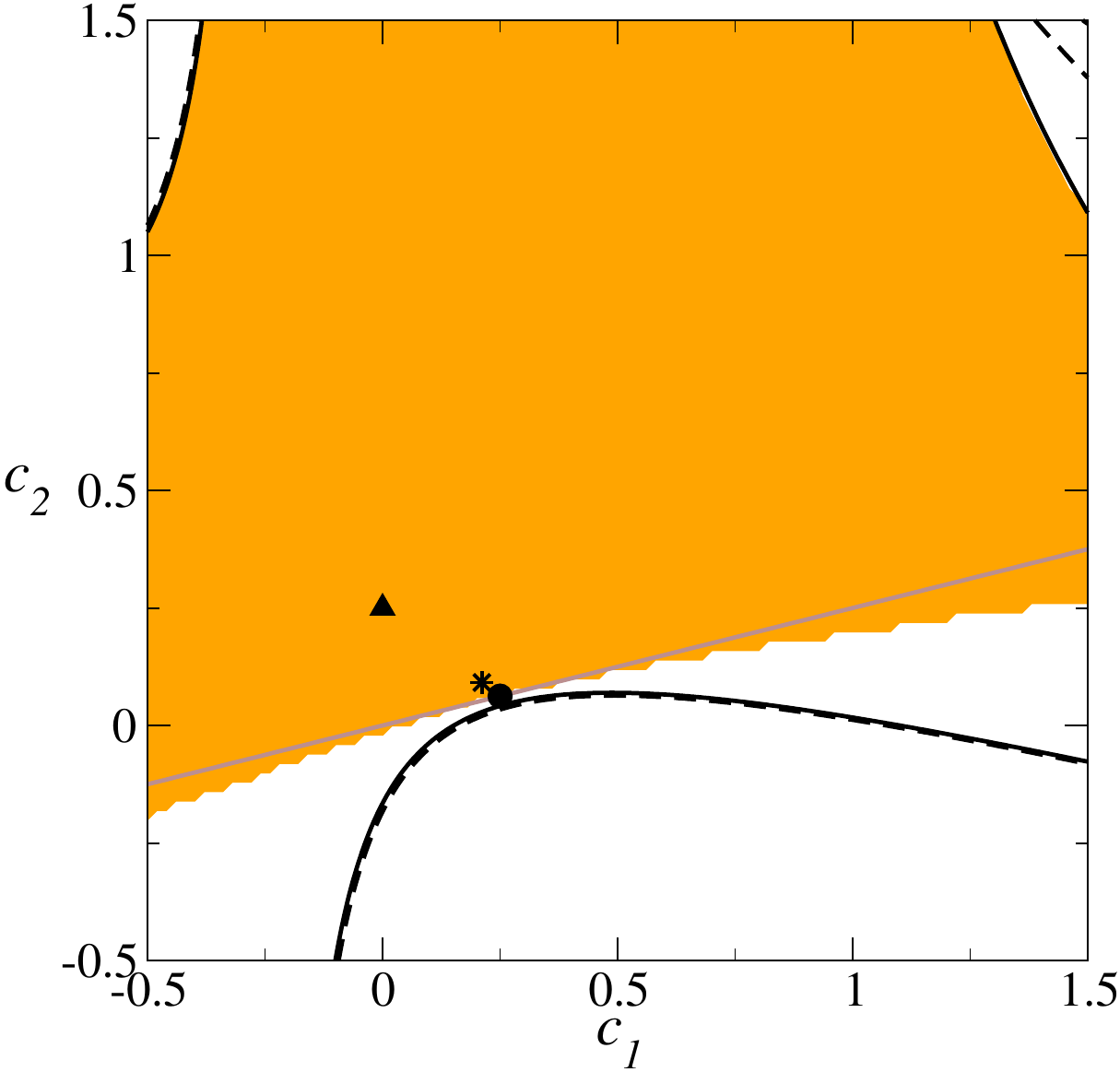}
	\hspace{0.05cm}
	\includegraphics[width=0.49\textwidth]{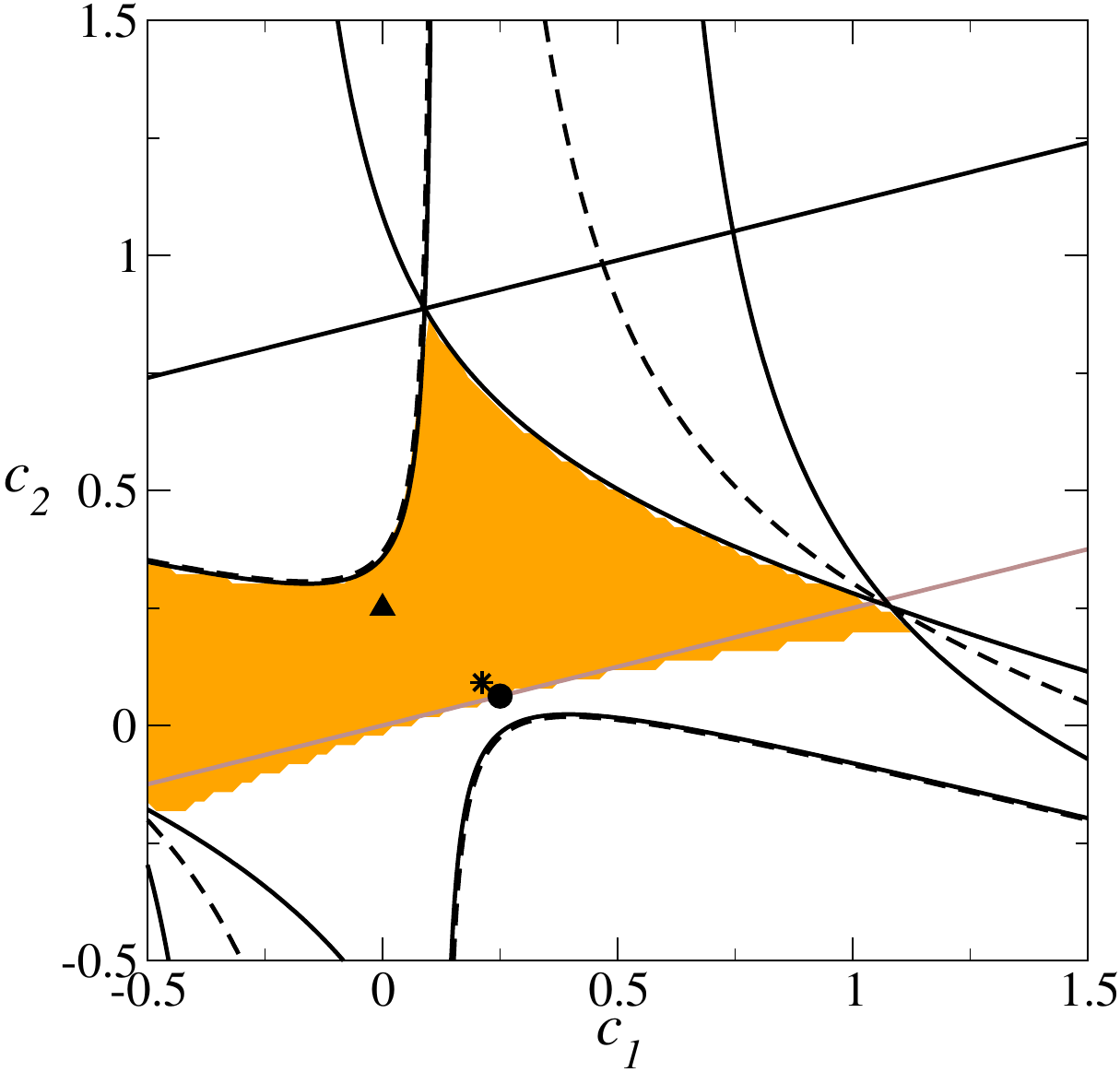}
	\\ \vspace{0.14cm}
	\includegraphics[width=0.49\textwidth]{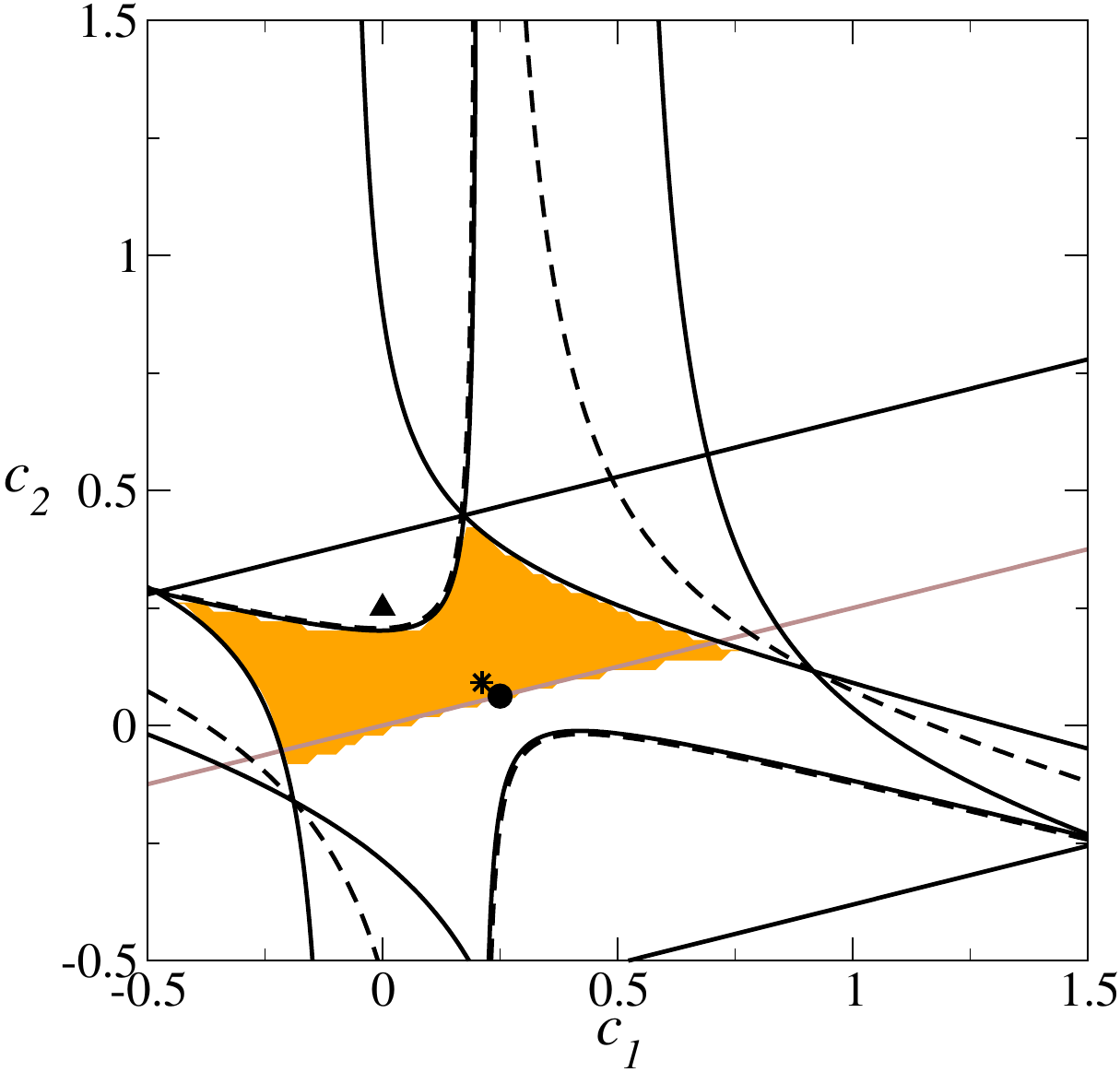}
	\hspace{0.05cm}
	\includegraphics[width=0.49\textwidth]{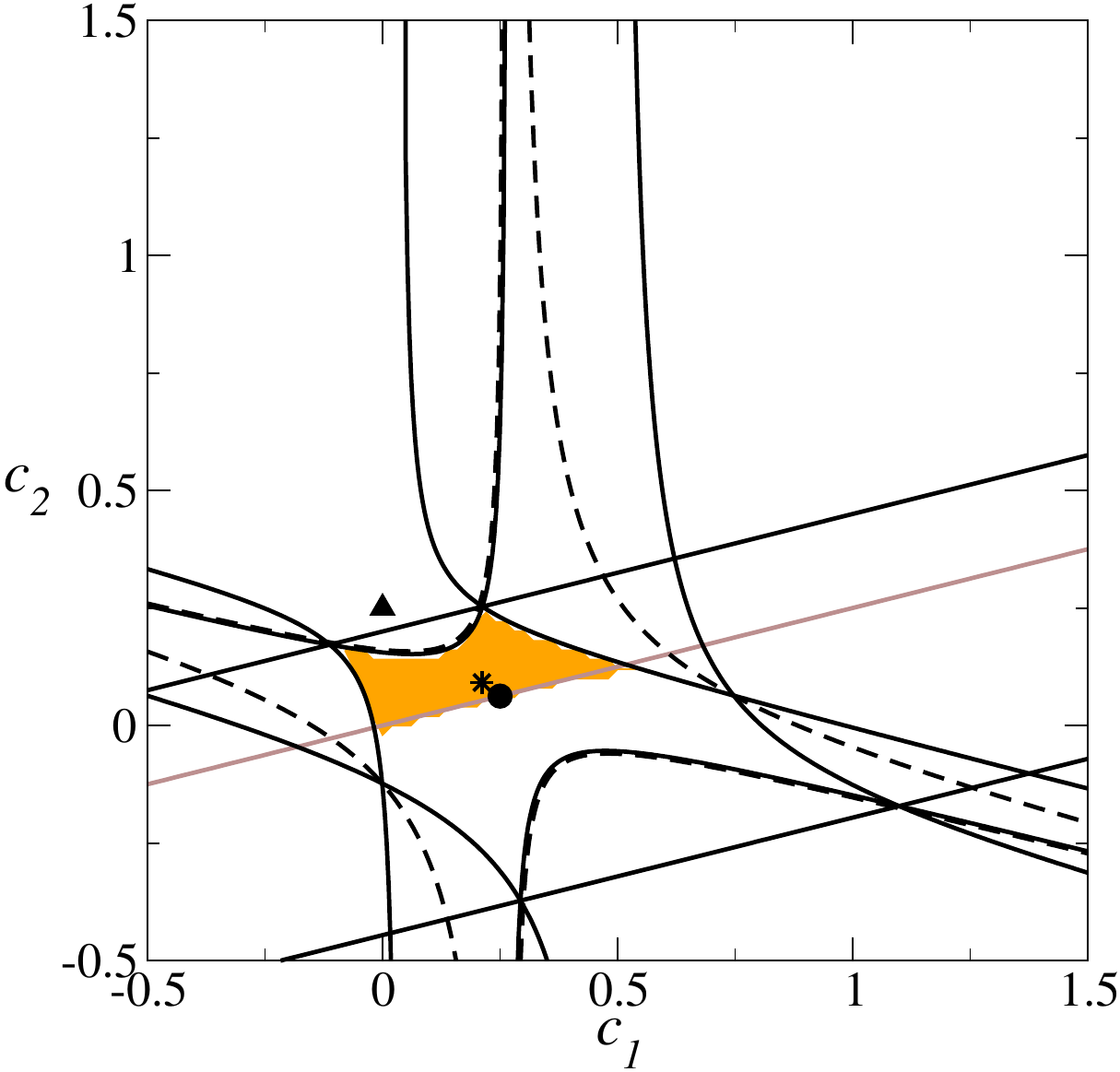}
	\caption{Dependence of the numerically determined stability region (orange 
	area) on the $(\ce_1, \ce_2)$ coefficients for the evolution of $h$ using a 
	third-order PIRK method. The boundaries (solid lines) of the stability region 
	according to Eqs.~(\ref{e:numbound3a})-(\ref{e:numbound3d}) for 
	$\bar{x} = 5.322$ (black lines) and $\bar{x}=0$ (brown line), and the 
	boundaries of the region $|\det(M_3)| \leq 1$, which partially coincide with 
	the stability boundaries, are plotted. The values for the optimal third-order 
	PIRK methods, $(\ce_1, \ce_2)=(1/4, 1/16)$  (PIRK3a, black circle) and 
	$(\ce_1, \ce_2)=((3 - \sqrt{3})/6, (-1 + \sqrt{3})/8)$ (PIRK3b, star symbol), and
	the third-order ERK scheme (black triangle), are also plotted.}
\label{fig:3}
\end{center}
\end{figure}
We have performed the same numerical stability analysis for the third-order 
PIRK method as in Sect.~\ref{sec:num2}. The numerically computed stability 
regions depending on $(\ce_1,\ce_2)$ are shown in Fig.~\ref{fig:3} for 
different CFL factors. As in the second-order method, the stability regions 
shrinks for increasing CFL factors. We have performed a $\chi^2$ minimization 
to determine the value of $\bar{x}$ which fits better the boundaries of the 
stability region with the expressions 
(\ref{e:numbound3a})-(\ref{e:numbound3d}). We obtain $\bar{x} = 5.322$ 
($\chi^2 = 109.0$ for $255$ degrees of freedom). For the numerical setup 
considered, we estimate $\bar{x}_{\rm min} = 6.945 \times 10^{-7} \ll 1$, so we
consider as a boundary the condition~(\ref{e:numbound3d}) setting $x=0$ (brown 
solid line in Fig.~\ref{fig:3}). The resulting boundaries (solid lines in 
Fig.~\ref{fig:3}) fit very well with the numerically estimated stability region
(orange area in Fig.~\ref{fig:3}). The condition $|\det{M_3}| \le 1$ (dashed 
lines in Fig~\ref{fig:3}) overestimate the stability region; however, the 
optimal values obtained in Sect.~\ref{sect:PIRK3} (black circle and star symbol
in Fig.~\ref{fig:3}) lay inside the stability region for all CFL factors 
studied. Since both values are very close, any of them could be used in the 
case of the wave equation for CFL factors close to unity. If we compare to the 
third-order IMEX-SSP3(4,3,3) scheme in~\cite{ParRus}, corresponding to 
$\ce_1 = 0.24169426078821$ and $\ce_2 = (1-3\ce_1)/4 = 0.06872930440884$, the 
stability properties are very similar since the coefficients are very close to 
the optimal PIRK values, and therefore the method is numerically stable for the
considered CFL factors.

The third-order ERK scheme corresponds to $(\ce_1,\ce_2) = (0, 1/4)$ (black 
triangle in Fig.~\ref{fig:3}). The eigenvalues of $M_3$ in this particular case
are
\begin{equation}
  e_\pm = 1 - \frac{x}{2} \pm \frac{1}{6} \sqrt{-x(x-6)^2}.
\end{equation}
The condition $|e_\pm|\le 1$ results in
\begin{equation}
	0 \le x \le 3.
\end{equation}
For the fitted value of $\bar{x}$, this condition restricts the stability of 
the third-order ERK scheme to CFL $< 0.751$. Accordingly, our numerical 
simulations show numerical stability for 0.5 and 0.7 CFL factors, but not for 
0.8 and 0.9. We conclude that, although the third-order ERK scheme is stable, 
the time step is more restricted as in the third-order PIRK method.

\subsection{Convergence}
\begin{figure}
\begin{center}
	\vspace{0.05cm} \includegraphics[width=0.9\textwidth]{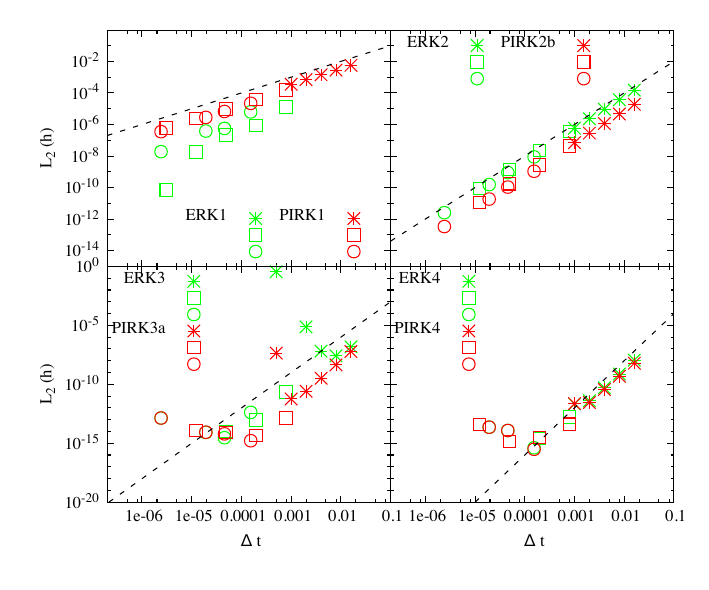}
	\caption{Numerical error estimation for series of 1D (stars), 2D (squares) 
	and 3D (circles) simulations of a scalar wave in spherical coordinates, as a 
	function of the time step, for first-order (upper left), second-order (upper 
	right), third-order (lower left) and fourth-order (lower right) methods. 
	Black-dashed lines corresponds to the expected order of convergence in each 
	case.}
\label{fig:4}
\end{center}
\end{figure}

We have studied the convergence properties of the PIRK methods by performing 
series of 1D-spherically symmetric, 2D-axisymmetric and 3D simulations. In each
one, we use reference models with resolutions $n_r = 50$, 
$(n_r, n_\theta) = (50, 16)$ and  $(n_r, n_\theta, n_\varphi) = (50, 8, 32)$, 
respectively. We increase the resolutions in all directions by the following 
factors: 2, 4, 8, 16 and 32 (1D); 2, 4, 8 (2D); 1.5, 2, 4 (3D). The $L_2$-norm 
of $h$ after $1/4$ period of oscillations, $t = \pi / \omega/2$, is used as an 
estimation of the error. To ensure that we measure the error due to the time 
discretization and not due to the spatial one, we choose the spatial order 
carefully. The time step in 1D, 2D and 3D simulations in spherical coordinates 
scales as $\Delta t_{\rm 1D} \sim \Delta r$, 
$\Delta t_{\rm 2D} \sim \Delta r \Delta \theta$, and
$\Delta t_{\rm 3D} \sim \Delta r \Delta \theta \Delta \varphi$, respectively, 
i.e., if we increase the resolution in all directions by a factor $2$, 
$\Delta t$ decreases in a factor $2$, $4$ and $8$, respectively. Therefore, to  
ensure that increasing resolution the spatial discretization errors decrease 
at least as fast as the time discretization errors, we choose fourth-order 
spatial discretization for the first-order methods, fourth (1D) and sixth-order
(2D, 3D) for the second-order methods, fourth (1D), eighth (2D) and tenth-order
(3D) for the third-order methods, and sixth (1D), tenth (2D) and twelfth-order 
(3D) for the fourth-order methods. In all cases, CFL~$=0.8$.

Fig.~\ref{fig:4} shows the error as a function of the time step. Independently 
of the dimensionality of the simulation, the error falls with decreasing time 
step as expected from the convergence order of the method used (dashed lines). 
In the case of third-order methods, the numerically obtained order seems 
closer to $4$ (dotted line); this is probably due to cancellations in the 
third-order time derivatives, for this particular test. Note that the error is 
limited by the machine accuracy (double precision was used) in the highest 
resolution simulations using the third and fourth-order methods, which produces
an increase of the error with increasing resolution.

\subsection{Cartesian coordinates}

According to the stability analysis of Sect.~\ref{sect:met-stab}, the results 
obtained for the wave equation should be independent of the coordinates used in
the discretization of the PDEs. In order to test this, we have performed 1D, 2D
and 3D numerical simulations of the scalar wave equation in Cartesian 
coordinates. Our computational domain is a box of size $1$. We impose periodic 
boundary conditions. In this case, Eq.~(\ref{e:wave}) has solutions of the form
\begin{equation}
	h (x,y,z, t) \sim  \sin (k_x x + \phi_x) \sin (k_y y + \phi_y) 
        \sin (k_z z + \phi_z) \cos{\omega t},
\label{eq:wavesol_car}
\end{equation}
being $k_x$, $k_y$ and $k_z$ the wave number in the different directions, 
$\phi_x$, $\phi_y$ and $\phi_z$ a phase, and 
$\omega = \sqrt{{k_x}^2+{k_y}^2+{k_z}^2}$ the oscillation frequency. The 
boundary conditions limit the possible choices of the wavenumbers to integer 
multiples of $2\pi$. For our tests we have experimented with $k_i = 2\pi$ and 
$k_i=4\pi$ ($i=x,y,z$). We have performed series of simulations comparing the 
PIRK methods with ERK methods at different resolutions (ranging from $100$ 
points to $1000$ points) and CFL numbers (ranging from $0.001$ to $1$).

For shake of brevity and to avoid repetition with the results in spherical 
coordinates, we will not describe in detail our numerical results using 
Cartesian coordinates. All the simulations that we have performed agree with 
the results in spherical coordinates in terms of stability properties of both 
PIRK and ERK methods, and in terms of convergence properties. Therefore, we can
conclude that the performance of this methods is independent of the coordinates
used in the spatial discretization as expected.

\section{Application 3: non-linear wave equation}
\label{sect:num:nl}

For our last test we consider the non-linear wave equation written as a 
first-order system in time,
\begin{eqnarray}
	\partial_t h &=& A, \nonumber \\
	\partial_t A &=& h_{xx} - V'(h),
\label{e:nlw}
\end{eqnarray}
where $V(h)$ is a nonlinear function of $h$. Eqs.~(\ref{e:nlw}) form a system 
of Hamiltonian partial differential equations whose Hamiltonian is,
\begin{equation}
\mathcal{H} = \int \left (\frac{1}{2} A^2 + \frac{1}{2} {h_x}^2 + V(h) \right) dx.
\end{equation}
Therefore, even if there are no general analytical solutions for this system, 
the Hamiltonian should be preserved during numerical evolutions and this can be
used to test our numerical methods in the nonlinear regime. We use a nonlinear 
potential of the form $V(h)=h^4/4$. For this choice of $V(h)$, if the initial 
data are $C^\infty$, then the solution at any time will be $C^\infty$ as well 
\cite{Strauss89}. This system of equations is also known to develop odd-mode 
instabilities for sufficiently large amplitudes of $h$.

We have followed the test setup of \cite{MacLachlan} and solved equations in 
the domain $x \in [0,2\pi]$ imposing periodic boundary conditions. Our initial  
data are $h_0(x) = a \cos(x) + 10^{-12} \sin(x)$ and $A_0 (x) =0$. The 
$\sin(x)$ term, used in \cite{MacLachlan}, is used to excite in a controlled 
way the odd-mode instabilities. Those instabilities appear for values of 
$a>\sim 1.85$ \cite{MacLachlan}. We have performed numerical simulations with 
$a=2$ using either $100$ or $500$ points and sixth-order discretization for the
spatial derivatives, which is sufficient to properly resolve the dynamics of 
the system and to keep the spatial-discretization error bellow the 
time-discretization one. The linear part of Eq.~(\ref{e:nlw}) corresponds to a 
wave-equation with propagation speed $1$. Therefore, a CFL condition for 
explicit methods can be computed and gives a maximum time-step of 
$\Delta t_{\rm max} = \Delta x$.

As a measure of the error in our numerical simulations we have used the time 
averaged $L_2$-norm of the relative variation of the Hamiltonian during the 
simulation,
\begin{equation}
	{\rm error}(\mathcal{H}) = \sqrt{\frac{1}{t_{end}}
\int_0^{t_{end}} \left ( \frac{\mathcal{H}_0 - \mathcal{H}}{\mathcal{H}_0} \right )^2 dt},
\end{equation}
being $t_{end}=2000$ ($\approx 318$ oscillation cycles) the end of the 
simulation, and $\mathcal{H}_0$ the initial value of the Hamiltonian. We 
consider a numerical evolution to be numerically unstable if this measure of 
the error exceeds $1$. For those cases the error typically grows exponentially 
with time.

\begin{figure}
\begin{center}
	\vspace{0.05cm} \includegraphics[width=0.95\textwidth]{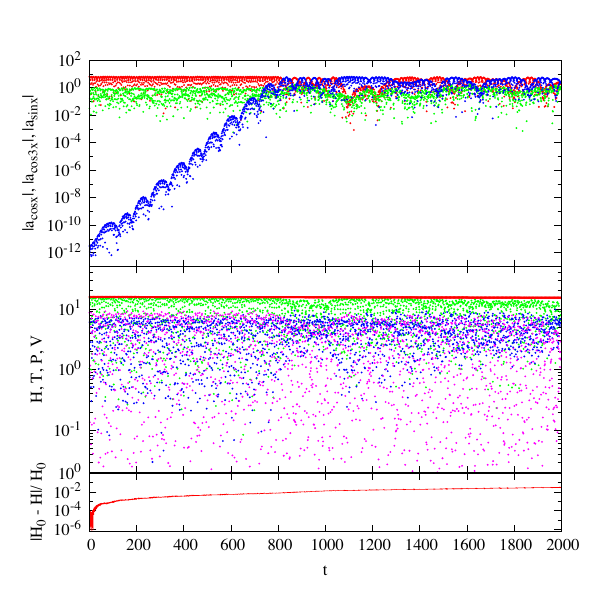}
	\caption{Time evolution of the nonlinear wave equation using the PIRK2a 
method, $\Delta t = 0.01$ and $500$ points. Upper panel shows the overlap 
integrals for the original perturbation ($\cos{x}$, red points), the first 
overtone ($\cos{3x}$, green points) and the odd-mode instability ($\sin{x}$, 
blue points). Middle panel shows the Hamiltonian, $\mathcal{H}$ (red points), 
and the different contributions to the integral: $\mathcal{T}$ (green points), 
$\mathcal{P}$ (blue points) and $\mathcal{V}$ (violet points). Lower panel 
shows the relative difference of $\mathcal{H}$ and its initial value 
$\mathcal{H}_0$.} 
\label{fig:nlw1}
\end{center}
\end{figure}

As an example for the typical evolution of a numerically stable simulation with
$500$ points and $\Delta t = 0.01 = 0.8 \Delta t_{\rm max}$, the middle panel 
of Fig.~\ref{fig:nlw1} shows the time evolution of the Hamiltonian 
$\mathcal{H}$, which stays constant within a few percent (lower panel). We also
plot separately the different contributions to the Hamiltonian:
\begin{eqnarray}
\mathcal{T} &\equiv& \int  \frac{1}{2} A^2  dx, \\
\mathcal{P} &\equiv& \int   \frac{1}{2} {h_x}^2 dx, \\
\mathcal{V} &\equiv& \int   V(h)  dx.
\end{eqnarray}
The contribution to the system of the nonlinear potential, $\mathcal{V}$, is 
non negligible and of the order of $\mathcal{P}$. Therefore, the evolution of 
the system is genuinely nonlinear. We have also computed the overlap integrals 
\begin{equation}
a_{f(x)} \equiv \int h(x) f(x) dx,
\end{equation}
using as trial functions the initial perturbation, $f(x)=\cos(x)$, an overtone 
$f(x)=\cos(3x)$ and an odd-mode, $f(x)=\sin(x)$, which is excited by the 
nonlinear interaction. The upper panel of Fig.~\ref{fig:nlw1} shows the time 
evolution of the time integrals, which are comparable to the results shown 
in~\cite{MacLachlan}.

\begin{figure}
\begin{center}
	\vspace{0.05cm} \includegraphics[width=0.95\textwidth]{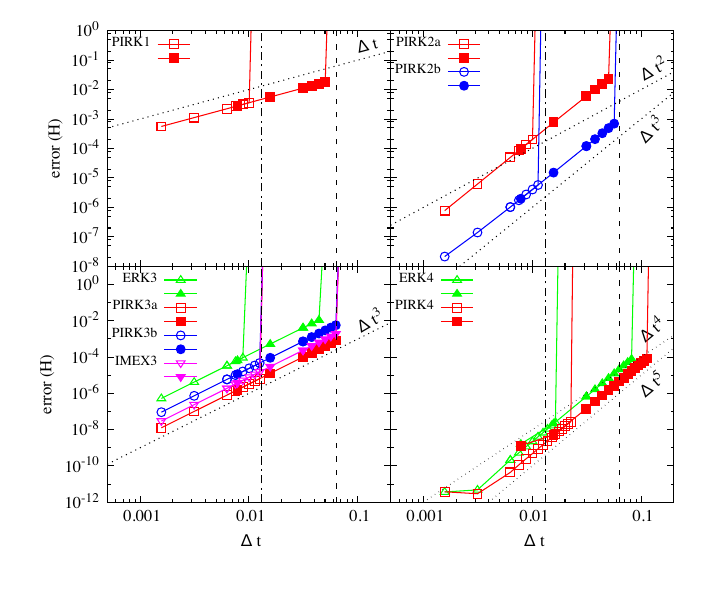}
	\caption{Time averaged $L_2$-norm of the relative change of the Hamiltonian 
at the end of the simulation ($t=2000$) in the evolution of the nonlinear wave 
equation, as a function of time-step $\Delta t$. Results for first (upper left 
panel), second (upper right), third (lower left) and fourth-order (lower right)
methods are plotted for the case with $100$ points (filled symbols) and $500$ 
points (open symbols). Dotted lines show difference power dependence on 
$\Delta t$, for reference. Vertical lines show the CFL condition for the case 
with $100$ (dashed lines) and $500$ points (dot-dashed lines).} 
\label{fig:nlw2}
\end{center}
\end{figure}

Fig.~\ref{fig:nlw2} shows the convergence and stability of the PIRK methods in 
comparison with ERK methods. For this purpose we performed simulations with CFL
factors ($\Delta t / \Delta t_{\rm max}$) ranging from $0.125$ to $2$. In 
almost all cases we were able to find a limiting CFL factor, within the range, 
above which numerical instabilities develop (${\rm error}(\mathcal{H})>>1$); 
their values are shown in table~\ref{tab:nlw:cflmax}. The exception are the 
first and second-order ERK methods for which all simulations developed 
numerical instabilities; this is consistent with the unconditionally unstable 
nature of those methods for wave-like equations, as seen in 
Sect.~\ref{sect:met-stab}. In the third and fourth-order cases, the PIRK 
methods were stable for larger CFL numbers that ERK methods, by a factor 
$\sim1.4$. Note that the maximum CFL factor for both resolutions ($100$ and 
$500$) is the same, but it corresponds to a different maximum time-step in each
case.

\begin{table}
\caption{Maximum CFL in numerical simulations of the non-linear wave equation, 
for which no numerical instabilities developed, depending on the numerical method
and the number of points used. In cases marked with "$-$" all numerical simulations developed
numerical instabilities.}
\label{tab:nlw:cflmax}
\centering
\begin{tabular}{|cc||cc|}
\hline
       &        & number of points & \\
       &        & $100$ & $500$ \\ \hline \hline
       & ERK1 	& -	  & -	  \\
			 & PIRK1  & 0.8 & 0.8 \\
			 & ERK2 	& -   & -   \\
			 & PIRK2a & 0.8 & 0.8 \\
			 & PIRK2b & 0.9 & 0.9 \\
method & ERK3   & 0.7 & 0.7 \\
			 & PIRK3a & 1.0 & 1.0 \\
			 & PIRK3b & 1.0 & 1.0 \\
			 & IMEX3  & 1.0 & 1.0 \\
			 & ERK4 	& 1.3 & 1.3 \\
			 & PIRK4	& 1.8 & 1.8 \\ \hline
\end{tabular}
\end{table}

The results in Fig.~\ref{fig:nlw2} show that for first and third-order methods 
the numerical error behaves as $\propto \Delta t$ and $\propto \Delta t^3$, 
respectively, as expected. For second and fourth-order methods the numerical 
error decays faster with decreasing $\Delta t$ than expected, very close to 
$\propto \Delta t^3$ and $\propto \Delta t^5$, respectively. This behaviour is 
probably due to the particular symmetries of the problem at hand, which produce
cancellations of even powers of $\Delta t$ in the expansion of the error in 
terms of the time-step. In all cases the error achieved by the PIRK methods 
were smaller than that in the ERK methods by a factor $\sim 10-100$ and 
$\sim 10$ for the third and fourth-order cases, respectively. The behaviour of 
the IMEX-SSP3(4,3,3) (IMEX3 here) was similar to the PIRK methods in all cases 
regarding both the maximum CFL and the error.

\section{Conclusions}
\label{sect:con}

In this work we present a new class of Runge-Kutta methods, the PIRK schemes, 
tailored to solve wave-like systems of non-linear PDEs. Methods up to 
fourth-order of convergence have been derived. The new methods are partially 
implicit in the sense that a proper subset of the equations of the system 
contains some terms which are treated implicitly. Optimal SSP ERK methods are 
recovered when implicitly treated parts are neglected. Although the implicit 
parts confer stability to the system, no analytical or numerical inversion of 
any operator is required, and the computational costs of the derived PIRK 
methods are comparable to those of the ERK ones.

To derive the new methods we first relate the coefficients of a $s$-stages PIRK
method to a $s+1$-stages IMEX one, and then we impose the order conditions to 
find the relations between the different coefficients of the method. The 
remaining free coefficients in each case are chosen according to stability 
criteria. For this purpose we analyse the determinant of the matrix which 
updates values from one time-step to the next one. Although this condition is 
less restrictive than the corresponding bound on the eigenvalues, which is the 
definition of the linear stability, we show, according to the numerical 
simulations performed, that the coefficients lead to numerically stable 
integrations.

We have successfully applied the derived PIRK methods to several cases: the 
evolution of a system of ODEs, the evolution of a scalar wave-equation and the 
evolution of a non-linear scalar wave-equation. We have studied the numerical 
stability of the systems and checked the convergence order of the PIRK methods 
in all the simulations, which were in agreement with the expected ones.

In the case of the scalar wave-equation, for first, second and third-order 
methods, we have studied the numerical stability of the system depending on the
free coefficients of the method and the CFL factor. This kind of analysis was 
numerically not feasible for the fourth-order PIRK method because of its high 
computational cost of having to explore the 5-dimensional space of free 
parameters. We have been able to compare the stability region estimated 
numerically with that obtained from the eigenvalues of the system, with 
excellent agreement. In all cases our optimal choice for the coefficients lays 
within the stability region, which shrinks with increasing resolution. This 
result is interesting because, despite of the fact that we computed the 
coefficients using the less restrictive condition for the determinant, the 
result still seems to be optimal. We think this is encouraging, as this kind 
of analysis could be used to compute optimal schemes for higher-order methods 
to evolve in time wave-like equations, without the complexity of having to work
with the eigenvalues. 

Applied to wave-like equations, stable numerical evolutions with the PIRK 
methods were possible using time-steps larger than in the traditional ERK 
methods. Stability properties for first and second-order methods are also 
superior to the ERK methods, the latter ones resulting to be unconditionally 
unstable in both the linear stability analysis and the numerical simulations. 
The third and fourth-order ERK methods are stable, but with a lower maximum 
time-step achievable than in the corresponding third and fourth-order PIRK 
methods. We suspect that the success of ERK methods in other works applied to 
wave-like equations could be due to the wide spread use of 
numerical-dissipation terms (e.g.~\cite{KreOli}), which could avoid the growth 
of small-scale unstable modes, as well as a sufficient small time-step in order
to control the growth of the numerical errors. The PIRK methods do not need any
numerical-dissipation term to get stable evolutions nor suffer from such 
time-step restrictions. Compared to IMEX methods, the presented second and 
third-order PIRK methods showed similar stability and convergence properties. 
However, some work had to be done to adapt the IMEX-SSP2(2,2,2) and the 
IMEX-SSP3(4,3,3) of~\cite{ParRus} to eliminate unnecessary stages. 

\begin{figure}
\begin{center}
	\vspace{0.05cm} \includegraphics[width=0.99\textwidth]{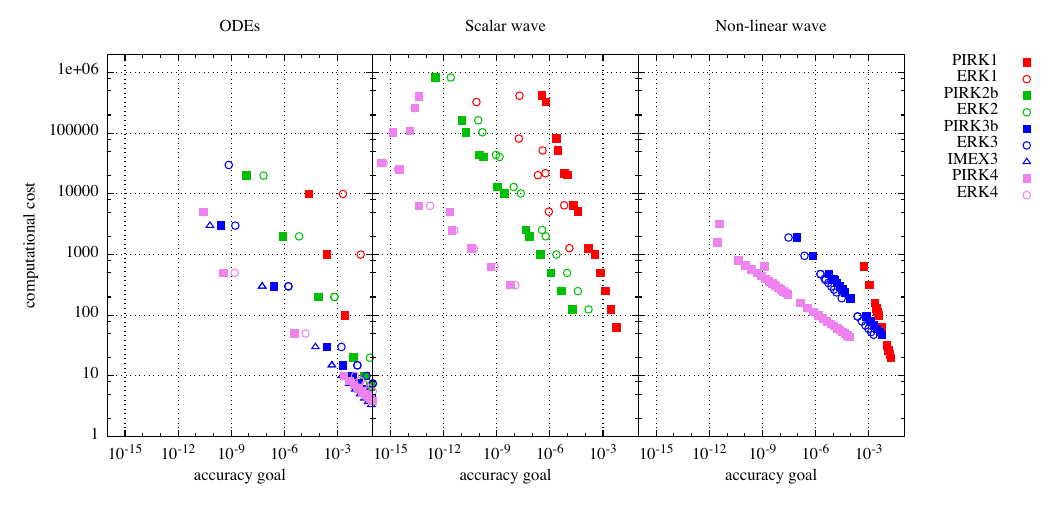}
	\caption{Computational cost, measured as the number of evaluations of the 
	$L_i$ operators per unit time, as a function of the error in the solution for
	the three tests presented in this work and different numerical methods.} 
\label{fig:cost}
\end{center}
\end{figure}

To quantify the advantage of the PIRK methods over the ERK ones, we can 
estimate the computational cost of each numerical simulation presented in 
Sects.~\ref{sect:num:ode}, \ref{sect:num:sph} and \ref{sect:num:nl} computing 
the number of evaluations of the $L_i$ operators per unit time, which is 
proportional to $s/\Delta t$, $s$ being the number of stages of the method. 
Since the bulk of the computational cost of most numerical applications comes 
from the evaluation of the $L_i$ sources, this is a good estimator of the 
computational cost in real applications. In Fig.~\ref{fig:cost} we plot the 
computational cost as a function of the numerical error measured in the 
simulation, for the different numerical methods and tests presented in this 
work. For a given accuracy goal, the plots tell us which is the numerical 
method with less computational cost. We have omitted in the plot those cases 
in which the error was decaying fastest than the expected order of convergence;
in those cases the error is particularly low due to symmetries of the 
particular test under consideration and cannot be used to estimate the error in
a general situation. In general, higher-order methods are less costly to obtain
the same numerical error. Comparing different methods, the PIRK schemes are 
less costly than the ERK ones of the same order. The IMEX schemes behave 
similarly to the PIRK methods of the same order. Taking into account of all the
methods tested, the PIRK4 scheme is the most efficient one in terms of 
computational cost.

An straightforward application of the PIRK methods is the numerical integration
of Einstein equations, which involves the solution of a hyperbolic system of 
non-linear equations with a wave-like structure. This system of equations is of
great interest for the numerical relativity community, in order to generate
gravitational wave templates for the upcoming gravitational wave detectors 
(LIGO, VIRGO and KAGRA). As a matter of fact, the PIRK methods presented in the
present work have already been applied to evolve Einstein equations \cite{CC12,
MonCC12, Baum13, Hilditch13, Nico14, Rekier15, Nico15a, Nico15b, BMM15, 
Rekier16, Nico16a, Nico16b}. These recent works represent the first 
multidimensional simulations of Einstein equations in spherical coordinates, 
that traditionally have been performed in Cartesian coordinates, and the use of
the PIRK methods was critical for the success of such simulations.

\appendix

\section{Stability conditions for second-order PIRK methods}
\label{a:PIRK2}
In this appendix we detail the derivation of the 
inequalities~(\ref{e:PIRK2boundc}) and (\ref{e:PIRK2boundx}). We would like to 
guarantee
\begin{equation}
	-4 \leq K_1 + \lambda \alpha_2 K_2 (1-2\ce_1+2\ce_2) 
+ \lambda^2 \alpha_2^2 (2\ce_2-\ce_1-2\ce_1\ce_2) \leq 4,
\label{e:boundapp2}
\end{equation}
for $\omega_i = 0,\pm 1$, $i=1,2$, and in the cases $|\lambda \alpha_2| \gg 1$,
$\lambda \alpha_2 \approx -1$ and $|\lambda \alpha_2| \ll 1$:
\begin{enumerate}
	\item Upper bound, $\omega_1 = \omega_2 = \pm 1$: $K_1=4$, $K_2=0$. In all the
cases we require $2\ce_2(1-\ce_1) - \ce_1 \leq 0$.
	\item Upper bound, $\omega_1 = - \omega_2 = -\pm 1$: $K_1=4$, $K_2=2$.
\subitem For $|\lambda \alpha_2| \ll 1$, we require $0 \leq 1 - 2\ce_1 + 2\ce_2$. 
This condition is sufficient for the cases $|\lambda \alpha_2| \gg 1$ and 
$\lambda \alpha_2 \approx -1$.
	\item Upper bound, $\omega_1 = \omega_2 = 0$: $K_1=1$, $K_2=1$. Previous 
derived conditions are sufficient for all the cases.
	\item Lower bound, $\omega_1 = \omega_2 = \pm 1$: $K_1=4$, $K_2=0$. In all the
cases we require $-8 \leq \lambda^2 \alpha_2^2 (2\ce_2-\ce_1-2\ce_1\ce_2)$. This 
condition will be satisfied by the following ones.
	\item Lower bound, $\omega_1 = - \omega_2 = -\pm 1$: $K_1=4$, $K_2=2$.
\subitem For $|\lambda \alpha_2| \ll 1$, we require 
$-4 \leq \lambda \alpha_2 (1-2\ce_+2\ce_2)$.
\subitem For $|\lambda \alpha_2| \gg 1$, previous derived conditions are 
sufficient.
\subitem For $\lambda \alpha_2 \approx -1$, we require 
$0 \leq 6 + 5\ce_1 - 6\ce_2 + 2\ce_1\ce_2$.
	\item Lower bound, $\omega_1 = \omega_2 = 0$: $K_1=1$, $K_2=1$.
\subitem For $|\lambda \alpha_2| \ll 1$, previous derived conditions are 
sufficient.
\subitem For $|\lambda \alpha_2| \gg 1$, we require 
$-5 \leq \lambda^2 \alpha_2^2 (2\ce_2-\ce_1-2\ce_1\ce_2)$.
\subitem For $\lambda \alpha_2 \approx -1$, we require 
$0 \leq 4 + \ce_1 - 2\ce_1\ce_2$.
\end{enumerate}

\section{Stability conditions for third-order PIRK methods}
\label{a:PIRK3}
In this appendix we detail the derivation of the 
inequalities~(\ref{e:PIRK3boundc}) and (\ref{e:PIRK3boundx}). We would like to 
guarantee 
\begin{eqnarray}
	-1 \leq &&\frac{K_1}{36} + \frac{\lambda \alpha_2 K_2 (-1 + \ce_1 - 4\ce_2)}{24} 
\nonumber \\
	&&+ \frac{\lambda^2 \alpha_2^2}{12} 
[\ce_1 - 4\ce_2 + (\mathrm{dex}-1)(4\ce_2 - \ce_1^2 - 4\ce_1 \ce_2))] \nonumber \\
	&&- \frac{\lambda^3 \alpha_2^3}{72} [-1 + 3(1 - 2\ce_1)(\ce_1 + 4\ce_2)] \leq 1.
\label{e:boundapp}
\end{eqnarray}
for $\omega_i = \pm 1$, $i=1,2$, and in the cases $|\lambda \alpha_2| \gg 1$, 
$\lambda \alpha_2 \approx -1$ and $|\lambda \alpha_2| \ll 1$:
\begin{enumerate}
	\item Upper bound, $\omega_1 = \omega_2 = 1$: $K_1=36$, $K_2=0$, 
$\mathrm{dex}=1$.
\subitem For $|\lambda \alpha_2| \ll 1$, we require $\ce_1 - 4\ce_2 \leq 0$.
\subitem For $|\lambda \alpha_2| \gg 1$, we require 
$-1 + 3(1 - 2\ce_1)(\ce_1 + 4\ce_2) \leq 0$.
\subitem For $\lambda \alpha_2 \approx -1$, previous derived conditions in this
appendix are sufficient.
	\item Upper bound, $\omega_1 = \omega_2 = -1$: $K_1=4$, $K_2=8$, 
$\mathrm{dex}=1$.
\subitem For $|\lambda \alpha_2| \ll 1$, we require 
$\lambda \alpha_2 (-1 + \ce_1 - 4\ce_2) \leq 8/3$.
\subitem For $|\lambda \alpha_2| \gg 1$, previous derived conditions are 
sufficient.
\subitem For $\lambda \alpha_2 \approx -1$, we require 
$0 \leq 41 + 18(\ce_1 - 4\ce_2) - 3(1 - 2\ce_1)(\ce_1 + 4\ce_2)$. Taking into account 
previous derived conditions, it is sufficient to require $-20/9 \leq \ce_1-4\ce_2$.
	\item Upper bound, $\omega_1 = 1 = -\omega_2$ or $\omega_2 = 1 = -\omega_1$: 
$K_1=-12$, $K_2=8$, $\mathrm{dex}=-1$.
\subitem For $|\lambda \alpha_2| \ll 1$, previous derived conditions are 
sufficient.
\subitem For $|\lambda \alpha_2| \gg 1$, previous derived conditions are 
sufficient.
\subitem For $\lambda \alpha_2 \approx -1$, we require 
$0 \leq 73 + 18\ce_1^2 - 180\ce_2 + 9\ce_1(3 + 8\ce_2)$.
	\item Lower bound, $\omega_1 = \omega_2 = 1$: $K_1=36$, $K_2=0$, 
$\mathrm{dex}=1$.
\subitem For $|\lambda \alpha_2| \ll 1$, we require 
$-24 \leq \lambda^2 \alpha_2^2 (\ce_1 - 4\ce_2)$.
\subitem For $|\lambda \alpha_2| \gg 1$, we require 
$\lambda^3 \alpha_2^3 [-1 + 3(1 - 2\ce_1)(\ce_1 + 4\ce_2)] \leq 144$. This condition 
will be satisfied by following ones.
\subitem For $\lambda \alpha_2 \approx -1$, we require 
$0 \leq 9\ce_1 - 12\ce_2 - 6\ce_1^2 - 24\ce_1 \ce_2 + 143$.
	\item Lower bound, $\omega_1 = \omega_2 = -1$: $K_1=4$, $K_2=8$, 
$\mathrm{dex}=1$.
\subitem For $|\lambda \alpha_2| \ll 1$, previous derived conditions are 
sufficient.
\subitem For $|\lambda \alpha_2| \gg 1$, we require 
$\lambda^3 \alpha_2^3 [-1 + 3(1 - 2\ce_1)(\ce_1 + 4\ce_2)] \leq 80$. This condition 
will be satisfied by following ones.
\subitem For $\lambda \alpha_2 \approx -1$, we require 
$0 \leq 103 - 15\ce_1 - 6\ce_1^2 + 84 \ce_2 - 24\ce_1 \ce_2$.
	\item Lower bound, $\omega_1 = 1 = -\omega_2$ or $\omega_2 = 1 = -\omega_1$:
$K_1=-12$, $K_2=8$, $\mathrm{dex}=-1$.
\subitem For $|\lambda \alpha_2| \ll 1$, previous derived conditions are 
sufficient.
\subitem For $|\lambda \alpha_2| \gg 1$, we require 
$\lambda^3 \alpha_2^3 [-1 + 3(1 - 2\ce_1)(\ce_1 + 4\ce_2)] \leq 48$.
\subitem For $\lambda \alpha_2 \approx -1$, we require 
$0 \leq 6\ce_1^2 - 15\ce_1 + 36\ce_2 + 24\ce_1 \ce_2 + 71$.
\end{enumerate}

\section*{Ackowledgements}
This work was supported by the European Research Council (Starting Independent 
Research Grant CAMAP-259276), the Spanish Ministerio de Econom\'ia y 
Competitividad (grants AYA2013-40979-P and AYA2015-66899-C2-1-P) and the 
Generalitat Valenciana (PROMETEO-II-2014-069).



\begin{thebibliography}{30}
%
\bibitem{Butcher}
Butcher, J.C.: Numerical Methods for Ordinary Differential Equations, 2nd ed. 
J. Wiley, Chichester (2008)
%
\bibitem{ARW}
Asher, U.M., Ruuth, S.J., Wetton, B.T.R.: Implicit-explicit methods for 
time-dependent PDE's. SIAM J. Num. Anal. 32, 797 (1995)
%
\bibitem{ARS}
Asher, U.M., Ruuth, S.J., Spiteri, R.J.: Implicit-explicit Runge-Kutta methods 
for time-dependent partial differential equations. Appl. Num. Math. 25, 151 
(1997)
%
\bibitem{Pareshi}
Pareshi, L.: Central differencing based numerical schemes for hyperbolic 
conservation laws with relaxation terms. SIAM J. Num. Anal. 39, 1395 (2001)
%
\bibitem{ParRus}
Pareschi, L., Russo, G.: Implicit-explicit Runge-Kutta methods and application 
to hyperbolic systems with relaxation. J. Sci. Comput. 25, 129 (2005)
%
\bibitem{deVogelaere} 
de Vogelaere, R.: Methods of integration which preserve the contact transformation property
of the Hamiltonian equations, Report No. 4, Dept. Math., Univ. of Notre Dame (1956)
%
\bibitem{Hairer}
Hairer, E., Wanner, G., Lubich, C.: Geometric numerical integration: Structure-preserving 
algorithms for ordinary differential equations, 2nd ed.
Springer-verlag, Berlin (2002)

\bibitem{Hairer2}
Hairer, E., Wanner: Solving ordinary differential equations II. Stiff and differentil-algebraic problems, 2nd ed.
Springer-verlag, Berlin (2002)

\bibitem{Soderlind}
S\"oderlind, G., Jay, L., Calvo, M.: Stiffness 1952:2012: Sixty years in search of a definition.
BIT Numer. Math. 55, 531-558 (2015)

\bibitem{Curtiss}
Curtiss, C.F. \& Hirschfelder, J.O.: Integration of stiff equations.
PNAS, 38, 235-243 (1952)

\bibitem{Vlach}
Vlach, J. and Singhal, K.: Computer methods for circuit analysis and design.

\bibitem{Rapaport}
Rapaport, D.C.: The art of molecular dynamics simulations, 2nd ed.
Cambridge University Press, Cambridge, UK (2004)

\bibitem{Yee}
Yee, K: "Numerical solution of initial boundary value problems involving Maxwell's equations in isotropic media". 
IEEE Transactions on Antennas and Propagation 14 (3): 302?307 (1966)


\bibitem{Baraff}
Baraff, D., Witkin, A.: Large steps in cloth simulation,
in SIGGRAPH, ACM, pp 43-54 (1998)

\bibitem{Asher}
Asher, U.~M.: Numerical methods for evolutionary differential equations.
SIAM, Philadelphia, USA (2008).

\bibitem{Niiranen}
Niiranen J.: Fast and accurate symmetric Euler algorithm for electromechanical simulations.
in Electrimacs proceedings vol. 1, pp 71-78 (1999)

\bibitem{Fang}
Fang, H., Lin, G., Zhang, R: The first-order symplectic Euler method for simulation of GPR wave propagation
in pavement structure
IEEE Transactions on geoscience and remote sensing, vol. 51. no 1, 93-98 (2013)

\bibitem{Holland}
Holland, R: Implicit three-dimensional finite differencing of Maxwell?s equations.
IEEE Trans. Nucl. Sci., vol. NS-31, pp. 1322?1326 1(984)

\bibitem{Moxley}
Moxley III, F.I., Byrnes, T., Fujiwara, F., Weizhong, D.: A generalized finite-difference time-domain scheme for solving nonlinear Schr\"odinger equations.
Comp. Phys. Comm. 183, 2434-2440 (2012)

\bibitem{Pretorius}
Pretorius, F.: Numerical relativity using a generalized harmonic decomposition.
Class. Quant. Grav. 22, 425-451 (2005)


\bibitem{Baker}
Baker, J~G., Centrella, D.C., Koppitz, M, van Meter, J.: Gravitational-Wave Extraction from an Inspiraling Configuration of Merging Black Holes.
Phys. Rev. Lett., 96, 111102 (2006) 

\bibitem{Centrella}
Centrella, J., Baker, J.~G., Kelly, B.~J., van~Meter, J.~R.: Black-hole binaries, gravitational waves, and numerical relativity.
Rev. Mod. Phys. 82, 3069 (2010) 
%
\bibitem{KreOli}
Kreiss, H.-O., Oliger, J.: Methods for the Approximate Solution of Time 
Dependent Problems. World Meteorological Organization, New York (1973)

%
\bibitem{GoShu98}
Gottlieb, S., Shu, C.-W.: Total Variation Diminishing Runge-Kutta schemes. 
Math. Compt. 67, 73--85 (1998)
%
\bibitem{GoShu01}
Gottlieb, S., Shu, C.-W., Tadmor, E.: Strong-stability-preserving high order 
time discretization methods. SIAM Review 43, 89--112 (2001)
%
\bibitem{ShuOsher}
Shu, C.-W., Osher, S.: Efficient implementation of essentially non-oscillatory 
shock-capturing schemes. J. Comput. Phys. 77, 439 (1988)
%
\bibitem{Kraaijevanger}
Kraaijevanger, J.F.B.M.: Contractivity of Runge-Kutta methods. BIT 31, 482-528 (1991)
%
\bibitem{SpiteriRuuth02}
Spiteri, R.J., Ruuth, S.J.: A new class of optimal high-order strong-stability-preserving time discretization
methods. SIAM J. Numer. Anal. 40, 469--491 (2002)
%
\bibitem{Ruuth06}
Ruuth, S.J.: Global optimization of explicit strong-stability-preserving Runge-Kutta methods. Math.
Comput. 75, 183--207 (2006)
%
\bibitem{RuuSpi}
Ruuth, S.J., Spiteri, R.J.: High-order strong-stability-preserving Runge-Kutta 
methods with downwind-biased spatial discretizations. SIAM J. Numer. Anal. 42, 
974 (2004)
%
\bibitem{KenCar}
Kennedy, C.A., Carpenter, M.H.: Additive Runge-Kutta schemes for 
convection-diffusion-reaction equations. Appl. Numer. Math. 44, 139-181 (2003)
%
\bibitem{HV}
Hundsdorfer, W., Verwer, J.G.: Numerical Solution of Time-Dependent 
Advection-Diffusion-Reaction Equations. Springer-Verlag, Berlin (2003)
%
\bibitem{polLa}
Handbook of Mathematical Functions with Formulas, Graphs, and Mathematical 
Tables. Abramowitz, M., Stegun, I.A. (eds.), pp. 878--879, and 883. Dover, New 
York (1972)
%
\bibitem{Strauss89}
Strauss, W.A.: Nonlinear wave equations. 
AMS Regional Conference Series no. 73. AMS, Providence (1989)
%
\bibitem{MacLachlan}
MacLachlan, R.: Symplectic integration of Hamiltonian wave equations. 
Numer. Math., 66, 465--492 (1994)
%
\bibitem{CC12}
Cordero-Carri\'on, I., Cerd\'a-Dur\'an, P., Ib\'a\~nez, J.M.: Gravitational 
waves in dynamical spacetimes with matter content in the Fully Constrained 
Formulation. Phys. Rev. D 85, 044023 (2012)
%
\bibitem{MonCC12}
Montero, P., Cordero-Carri\'on, I.: BSSN equations in spherical coordinates 
without regularization: Vacuum and nonvacuum spherically symmetric spacetimes.
Phys. Rev. D 85, 124037 (2012)
%
\bibitem{Baum13}
Baumgarte, T.W., Montero, P., Cordero-Carri\'on, I., M\"uller, E.: Numerical 
relativity in spherical polar coordinates: Evolution calculations with the BSSN
formulation. Phys. Rev. D 87, 044026 (2013)
%
\bibitem{Hilditch13}
Hilditch, D., Baumgarte, T.W., Weyhausen, A., Dietrich, T., Bruegmann, B., 
Montero, P. M\"uller, E.: Collapse of Nonlinear Gravitational Waves in 
Moving-Puncture Coordinates. Phys. Rev. D 88, 103009 (2013)
%
\bibitem{Nico14}
Sanchis-Gual, N., Montero, P., Font, J.A., M\"uller, E., Baumgarte, T.W.: Fully
covariant and conformal formulation of the Z4 system in a reference-metric 
approach: comparison with the BSSN formulation in spherical symmetry. 
Phys. Rev. D 89, 104033 (2014)
%
\bibitem{Rekier15}
Rekier, J., Cordero-Carri\'on, I., F\"uzfa, A.: Fully relativistic nonlinear 
cosmological evolution in spherical symmetry using the BSSN formalism. 
Phys. Rev. D 91, 024025 (2015)
%
\bibitem{Nico15a}
Sanchis-Gual, N., Degollado, J.C., Montero, P., Font, J.A.: Quasi-stationary 
solutions of self-gravitating scalar fields around black holes. Phys. Rev. D 
91, 043005 (2015)
%
\bibitem{BMM15}
Baumgarte, T.W., Montero, P., M\"uller, E.: Numerical Relativity in Spherical 
Polar Coordinates: Off-center Simulations. Phys. Rev. D 91, 064035 (2015)
%
\bibitem{Nico15b}
Sanchis-Gual, N., Degollado, J.C., Montero, P., Font, J.A., Mewes, V.: 
Quasistationary solutions of self-gravitating scalar fields around collapsing 
stars. Phys. Rev. D 92, 083001 (2015)
%
\bibitem{Rekier16}
Rekier, J., F\"uzfa, A., Cordero-Carri\'on, I.: Nonlinear cosmological 
spherical collapse of quintessence. Phys. Rev. D 93, 043533 (2016)
%
\bibitem{Nico16a}
Sanchis-Gual, N., Degollado, J.C., Herdeiro, C., Font, J.A., Montero, P.: 
Dynamical formation of a Reissner-Nordstr\"om black hole with scalar hair in a 
cavity. Phys. Rev. D 94, 044061 (2016)
%
\bibitem{Nico16b}
Sanchis-Gual, N., Degollado, J.C., Izquierdo, P., Font, J.A., Montero, P.: 
Quasistationary solutions of scalar fields around accreting black holes. 
Phys. Rev. D 94, 043004 (2016)
 
\end{thebibliography}
\end{document}